\documentclass[11pt]{article}
\usepackage{amsmath}
\usepackage{amsfonts}
\usepackage{mathrsfs}
\usepackage{scalerel}
\usepackage{float}
\usepackage{slashed}
\usepackage{cite}
 \usepackage{tabu}
 \usepackage{soul}

\usepackage[colorlinks=true,
linkcolor=Blue, 
citecolor=Blue,
filecolor=Blue,
urlcolor=Blue,
linktoc=page, %%%
pdfstartview=FitV,
bookmarksopen=true]{hyperref}

\usepackage[dvipsnames]{xcolor}
\usepackage{eso-pic}% http://ctan.org/pkg/eso-pic
\usepackage{datetime}

\topmargin - 0.8 in

\def\d{\delta}

\def\s{\sigma}
\def\l{\lambda}

\newcommand{\bi}{\begin{itemize}}
\newcommand{\ei}{\end{itemize}}

\newcommand{\be}{\begin{equation}}
\newcommand{\ee}{\end{equation}}

\newcommand{\bea}{\begin{eqnarray}}
\newcommand{\eea}{\end{eqnarray}}

\numberwithin{equation}{section}
\makeatletter
\renewcommand{\@seccntformat}[1]{%
  \csname the#1\endcsname.\ }
\makeatother

\title{
\vskip-1mm
Non-linear asymptotic symmetries in warped $AdS_3$ holography\vskip5mm}

\author{ Silvia Georgescu \vspace{1mm}\\
\\\vspace{1mm}
\emph{\small 
Department of Mathematics, King’s College London, Strand, London, WC2R 2LS, United Kingdom} \\  \vspace{1mm}
}

\date{}

\begin{document}

\maketitle

%\AddToShipoutPictureBG*{
 % \AtPageUpperLeft{
%    \hspace{\paperwidth}
 %   \raisebox{-\baselineskip}{
%      \makebox[0pt][r]{\today ~~ \currenttime ~~~~~}
%}}}

\abstract{
\vskip2mm

\noindent Warped AdS$_3$ backgrounds provide set-ups to study holography beyond AdS and, in particular, holography for near-extremal Kerr black holes. A certain $U(1)$ charged warped BTZ background supported by pure NS-NS flux was constructed in string theory in \cite{Apolo:2021wcn}. While older works found, in absence of $U(1)$ charges, that the warped black holes' thermal entropy obeys a Cardy formula, the addition of $U(1)$ charges in \cite{Apolo:2021wcn} leads to the universal entropy formula in a $J\bar{T}$-deformed CFT and not to the charged Cardy formula. In this article, we explore further the implications of this result for warped $AdS_3$ holography. We compute the asymptotic symmetries of the warped BTZ background of \cite{Apolo:2021wcn} and obtain an infinite-dimensional non-linear Poisson algebra that can be linearized, after a non-linear redefinition of generators, to two commuting copies of the $(Virasoro\times U(1)Kac-Moody)$ algebra. The algebra matches the symmetry algebra of a symmetric product orbifold of $J\bar{T}$-deformed CFTs. We contrast the results with those for a similar warped BTZ background supported by both NS-NS and RR flux.}

\tableofcontents

\section{Introduction}
Understanding holography beyond the AdS/CFT correspondence is a very challenging open problem. From a top-down perspective, two interconnected reasons are the scarcity of tractable non-AdS string theoretical constructions and the difficulties in understanding non-standard strongly-coupled non-gravitational theories (some examples are \cite{Seiberg:1999vs,Maldacena:1999mh,Seiberg:1997zk,Losev:1997hx,Aharony:1998ub,Seiberg:2000ms,Harmark:2000wv,Gopakumar:2000ep,Bergman:2000cw}) which are expected to be dual to some of these non-AdS backgrounds.

A particularly interesting non-AdS background is the near-horizon geometry of (near-)extremal Kerr black holes (NHEK) \footnote{It was understood recently that the classical NHEK geometry cannot be trusted because of significant quantum corrections \cite{Iliesiu:2020qvm,Heydeman:2020hhw}. Our comments here apply as well for the near-extremal case that should be considered instead, but we prefer in the introduction to refer to the NHEK case for historical reasons.}. Motivated by the success of the holographic microscopic description of asymptotically AdS black holes, it is natural to ask whether such a description is available also for highly rotating Kerr black holes, which are a good theoretical model for black holes in our universe. The NHEK geometry has $SL(2,\mathbb{R})\times U(1)$ isometries. In \cite{Guica:2008mu}, it was shown that the $U(1)$ symmetry is enhanced asymptotically to one copy of the Virasoro algebra and that the black hole entropy obeys a Cardy-like formula with respect to its central charge. Based on these results, it was initially proposed that the holographic description of the NHEK geometry is given by a chiral two-dimensional CFT (the ``Kerr/CFT" correspondence) \cite{Guica:2008mu,Compere:2012jk}. However, despite the presence of Virasoro symmetry, computations of correlation functions \cite{Bredberg:2009pv,Becker:2010jj}, as well as constructions in string theory \cite{El-Showk:2011euy}, showed that the dual field theory is non-local and should be obtained as an irrelevant Lorentz-breaking deformation of a two-dimensional CFT, which starts with conformal dimension $(1,2)$ and leads to a UV-complete theory. Such deformation is of particular interest also from a field-theoretical point of view, as deformations of this type typically do not lead to UV-complete theories.

A simpler three-dimensional set-up that captures the main features of the NHEK geometry is the warped AdS$_3$ factor,  obtained by fixing the polar angle in NHEK. Warped AdS$_3$ backgrounds, generally defined as $U(1)$ fibrations over AdS$_2$, are natural generalizations of AdS$_3$ backgrounds, which preserve $ SL(2,\mathbb{R})\times U(1)$ isometries. They appear universally in the near-horizon region of extremal black holes in any number of dimensions, can be realized in string theory, and some have been related to dipole deformations \cite{Song:2011sr,El-Showk:2011euy}, which provide examples of non-local field theories. In particular, the fact that they can be obtained from top-down constructions in string theory makes warped AdS$_3$ holography a promising toy model for Kerr holography and, more generally, for understanding holography beyond AdS/CFT.

In this article, we focus on a particular warped AdS$_3$ background supported by pure NS-NS flux, which can be constructed in string theory and generalized to charged warped BTZ backgrounds \cite{Apolo:2021wcn},\footnote{We want to emphasize that, reduced to three dimensions, these backgrounds are not  quotients of warped AdS$_3$, which appear in bottom-up approaches to warped AdS$_3$ holography. Nevertheless, they are black hole solutions obtained by deforming BTZ. In \cite{Apolo:2021wcn}, they are referred to as ``TsT black holes".} via a TsT (T-duality, shift, T-duality) transformation of type IIB string theory on charged $BTZ\times S^3\times T^4$. An additional shift of the dilaton produced by the naive TsT transformation is necessary in order for the electric charge of the backgrounds to be quantized \cite{Apolo:2021wcn}.

One reason why these particular backgrounds are promising set-ups for studying warped AdS$_3$ holography is precisely the way they are constructed as deformations of AdS$_3$. As explained in detail in \cite{Apolo:2019zai}, when a worldsheet description is available for the undeformed theory, as it is in this case, the TsT transformation corresponds to an exactly marginal deformation of the string worldsheet theory by the antisymmetric product of two Noether currents. Hence, the advantage of these warped backgrounds supported by pure NS-NS flux is that one can analyze the deformed theory in string perturbation theory. Moreover, the deformed worldsheet theory can be mapped to the undeformed one with non-local twisted boundary conditions for the fields \cite{Alday:2005ww}. This observation was used, for example, in order to compute the worldsheet spectrum in the deformed theory \cite{Apolo:2021wcn}.

On the non-gravitational side, the TsT transformation should correspond to an irrelevant deformation starting with conformal dimension $(1,2)$ of the CFT$_2$ dual to the undeformed string theory. What is known regarding this irrelevant deformation is that on the long string sector of type IIB string theory on the undeformed backgrounds \eqref{undefbackgr}, which is well-described by a ``dual" symmetric product orbifold CFT\footnote{The long string sector is not decoupled from the rest of the theory, so this is not an usual holographic duality.}, it acts like a $J\bar{T}$ deformation of the seed CFT, also called ``single-trace $J\bar{T}$" \cite{Apolo:2018qpq,Chakraborty:2018vja}.\footnote{In \cite{Apolo:2018qpq,Chakraborty:2018vja}, the long string spectrum was computed in the deformed worldsheet theory and was shown to reproduce the single-trace $J\bar{T}$ spectrum.} A relation between the warped backgrounds obtained by TsT and single-trace $J\bar{T}$ is indicated also by thermodynamics. In \cite{Apolo:2021wcn} it was shown that the Bekenstein-Hawking entropy of the warped BTZ backgrounds takes the universal form of the entropy in single-trace $J\bar{T}$.\footnote{The status is very similar to that of the relation \cite{Giveon:2017nie} between the asymptotically linear dilaton background that arises in the near-horizon of the NS5-F1 system and the solvable irrelevant deformation called $T\bar{T}$ \cite{Smirnov:2016lqw,Cavaglia:2016oda}. Nevertheless, the advantage of the linear dilaton background is that it can be obtained directly from a decoupling limit in string theory, which indicates the non-gravitational dual as type IIB Little string theory compactified to two dimensions.} We note that the additional shift in the dilaton obtained via TsT is crucial for the entropy matching.

The $J\bar{T}$ deformation \cite{Guica:2017lia} is a universal, irrelevant, Lorentz-breaking deformation of two-dimensional CFTs with a conserved $U(1)$ current, which leads to non-local, but UV-complete theories. As mentioned, these  features are shared by the non-gravitational theories in warped AdS$_3$ holography. Although the deformation breaks half of the conformal symmetry, it preserves an infinite-dimensional symmetry algebra  \cite{Guica:2020uhm,Guica:2020eab,Guica:2021pzy,Georgescu:2024ppd}. The most up-to-date analysis of the symmetries of $J\bar{T}$-deformed CFTs can be found in \cite{Georgescu:2024ppd}. The symmetry algebra consists of a non-linear modification of the $(Virasoro\times U(1) Kac-Moody)^2$ algebra of the undeformed CFT, which can be linearized to $(Virasoro\times U(1)Kac-Moody)^2$ by a non-linear combination of generators. We emphasize that, despite the CFT-like algebra in a specific basis, the corresponding symmetry transformations are highly non-local (see \cite{Georgescu:2024ppd} for an example). In \cite{Guica:2021fkv}, this symmetry algebra was used to define a set of operators which are analogues of primary operators in 2d CFTs and compute their correlation functions, which reproduce the main features of scattering off near-extremal Kerr.

Turning to holography, the non-gravitational dual of a $J\bar{T}$-deformed CFT consists in three-dimensional gravity coupled to one or two $U(1)$ Chern-Simons gauge fields with non-standard boundary conditions \cite{Bzowski:2018pcy,Georgescu:2024ppd}. In \cite{Georgescu:2024ppd}, the $J\bar{T}$ symmetry algebra was reproduced holographically from the bulk. For the purpose of understanding holography beyond AdS, a single-trace version of the deformation was introduced for symmetric product orbifold CFTs. In the field theoretical analysis of \cite{Chakraborty:2023wel} it was shown that single-trace $J\bar{T}$ inherits the infinite-dimensional symmetry algebra of the seed theory.

Highly-tractable irrelevant deformations are rare. The fact that the $J\bar{T}$ deformation and its single-trace version (same for the $T\bar{T}$ deformation) are under control can be seen as a consequence of the high amount of symmetry constraining them. Coming back to the warped backgrounds of \cite{Apolo:2021wcn}, we address the question whether such amount of symmetry is also shared by their holographic duals, which resemble single-trace $J\bar{T}$. In this paper, we compute the asymptotic symmetry algebra of these backgrounds and indeed obtain an infinite dimensional non-linear Poisson algebra, which, moreover, matches perfectly the symmetry algebra of a symmetric product orbifold of $J\bar{T}$-deformed CFTs. In light of the fact that the two are not holographic dual, the matching requires further investigation.

In our analysis, we use the method outlined in \cite{Georgescu:2022iyx} for the case of the asymptotically linear dilaton background to select a set of boundary conditions that lead to a consistent phase space, which we construct perturbatively around the warped BTZ background. Using the same non-linear change of basis of generators as in \cite{Guica:2021pzy}, the resulting symmetry algebra is brought to the form of two commuting copies of the $(Virasoro\times U(1) Kac-Moody)$ algebra. The interplay between the two bases of generators reconciles the non-locality of general warped AdS$_3$ holographic duals with the enhancement of global $U(1)$ symmetries to Virasoro, suggesting a similar scenario in the case of (near-)extremal Kerr.

Non-linear asymptotic symmetry algebras are not common. In our case, the origin of non-linearity is the presence of charge-dependent parameters in the allowed transformations in phase space. In the holographic dual, such symmetry transformations reflect the non-local nature of the theory. These features, which were not observed before in warped AdS$_3$ holography, are visible only in the presence of the $U(1)$ charges, which were not considered in early warped AdS$_3$ string constructions, such as \cite{Song:2011sr}. If we turn off the $U(1)$ charges in our analysis, we obtain two usual commuting copies of the Virasoro algebra. This remark leads to the question of whether adding $U(1)$ charges to other warped AdS$_3$ set-ups leads to similar features, which can then be conjectured to be universal in warped AdS$_3$ holography. We looked at one particular example for which the answer seems\footnote{It is possible that the $U(1)$ charges we considered are not the ones corresponding to the current entering the $J\bar{T}$-like deformation. If this is the case, it is not surprising that the symmetry transformations that we obtain are CFT-like. Hence, it would be worth investigating whether there exist other $U(1)$ charges that can be turned on in this set-up which would lead to symmetry transformations similar to the ones for the pure NS-NS set-up of \cite{Apolo:2021wcn}. The author thanks Monica Guica for this comment.} to be no: TsT on the $BTZ\times S^3\times T^4$ background with pure RR flux, which appears in the near-horizon of the D1-D5 system. In this example, the non-locality inherent to warped AdS$_3$ duals is not visible from the asymptotic symmetries. Therefore, different from AdS/CFT, symmetry algebras do not seem to be universal in warped AdS$_3$ holography\footnote{Of course, asymptotic symmetry algebras depend on boundary conditions, but in the two stringy cases studied in this article the boundary conditions are singled out by the same consistency conditions.}.

The paper is organized as follows. In section \ref{section2:review}, we review the charged warped BTZ backgrounds of \cite{Apolo:2021wcn}, obtained by restricting their analysis to the particular case of a $J\bar{T}$-like deformation. In section \ref{section3:asysym} we construct perturbatively the phase space and compute  the asymptotic symmetry algebra. We discuss the results and suggest some future directions in section \ref{section4:discussion}. We include appendix \ref{appA:covphsp} to show details of the covariant phase space formalism, appendix \ref{appB:JTbar} to compare with the $J\bar{T}$ holographic analysis of \cite{Georgescu:2024ppd} and  appendix \ref{appC:details} for the details of the computations of allowed symmetry transformations. We also include appendix \ref{appD:dipole} for an analysis of another warped background for which the $J\bar{T}$ features seem to be absent.

\section{Review of the warped BTZ backgrounds}
\label{section2:review}
In \cite{Apolo:2021wcn}, various string backgrounds were generated using TsT (T-duality, shift, T-duality) transformations on the charged $BTZ\times S^3\times T^4$ background supported by pure NS-NS flux. These deformations map to exactly marginal deformations on the string worldsheet. Among them, a particular case is that of a family of warped BTZ backgrounds which were shown to reproduce the entropy formula which is universal for a particular deformation of 2d CFTs, the $J\bar{T}$ deformation. In this section, we review these backgrounds and their relation to the $J\bar{T}$ deformation of two-dimensional CFTs.

\subsection{The backgrounds}
The starting point are the following $U(1)$ charged $BTZ\times S^3 \times T^4$ backgrounds, supported by pure NS-NS flux: 
\begin{align}\label{undefbackgr}
ds^2&=\frac{\ell^2 dr^2}{4(r^2-4T_u^2 T_v^2)}+\ell^2(T_u^2+Q_L^2)dU^2+\ell^2(T_v^2+Q_R^2)dV^2+\ell^2(r-2 Q_R Q_L)dUdV+\nonumber\\
&+\ell^2 d\Omega_3^2+\ell_4^2 dy^2+2\ell\ell_4(Q_L dU-Q_R dV)dy+ \ell^2_4\sum_{i=8}^{10}dy_i^2
\end{align}
\begin{align}
B&=\frac{\ell^2}{4}\cos\theta d\phi\wedge d\chi+\ell^2\bigg(\frac{r}{2}-Q_L Q_R\bigg)dV\wedge dU+\ell\ell_4 dy\wedge(Q_L dU+Q_R dV)\\
e^{2\Phi}&=\frac{k}{p}k_4^2
\end{align}
where $k=\ell^2/\ell_s^2$, $k_4=\ell_4^2/\ell_s^2$, $\ell$ is the scale of $AdS_3$ and $\ell_4$ is the scale of $T^4$. The null coordinates $U=\sigma+t,V=\sigma-t$ are identified as:
\begin{align}
U\sim U +2\pi \hspace{1cm}V\sim V+2\pi
\end{align}
We denoted by $d\Omega_3^2$ the metric of the unit $S^3$, which in Hopf coordinates is written as:
\begin{align}
d\Omega^2_3&=\frac{1}{4}(d\theta^2+d\phi^2+d\chi^2+2\cos\theta d\phi d\chi)
\end{align}
The coordinates on $T^4$ are denoted as $\{y,y_{8,9,10}\}$ and have periodicity  $2\pi$. The backgrounds are characterized by four constant parameters, $T_{u,v},Q_{L,R}$, which encode the left and right-moving energies and two $U(1)$ charges (their expressions are obtained by setting $\lambda=0$ in \eqref{globalcha}). The backgrounds also carry electric and magnetic charges, corresponding to the number $p$ of F1 strings and the number $k$ of NS5 branes that support the backgrounds, respectively.

In order to obtain the charged warped BTZ backgrounds, one first does a TsT transformation: T-duality on $y$, shift $V\rightarrow V + \frac{\lambda}{\ell\ell_4} y$, T-duality back on $y$. After the TsT, the electric charge of the background is given by:
\begin{align}
\frac{1}{(2\pi \ell_s)^6}\int_{S^3\times T^4} e^{-2\Phi} *H= p(1+\lambda Q_R)
\end{align}
The electric charge should still count the number of F1 strings that support the background, so the fact that it is not quantized (and equal to $p$) is problematic. As explained in \cite{Apolo:2021wcn}, this issue can be fixed by a constant shift of the dilaton, which renders the electric charge to be quantized. Since the background is supported by NS-NS flux only, after this modification of the dilaton, it will still be a solution of the type IIB supergravity equations of motion. The resulting background, with the shifted dilaton is, in string frame:
\begin{align}\label{backgrwarp}
ds^2&=\frac{\ell^2 dr^2}{4(r^2-4 T_u^2 T_v^2)}+\ell^2(T_u^2+Q_L^2)dU^2 +\frac{\ell^2(T_v^2+Q_R^2)}{(1+\lambda Q_R)^2+\lambda^2 T_v^2}dV^2+\frac{\ell^2_4 }{(1+\lambda Q_R)^2+\lambda^2 T_v^2}dy^2+\nonumber\\
&+2\ell^2\frac{\frac{r}{2}-Q_L Q_R -\lambda Q_L(T_v^2+Q_R^2)}{(1+\lambda Q_R)^2+\lambda^2 T_v^2}dUdV+\ell^2 d\Omega_3^2+2\ell \ell_4 \frac{\frac{r\lambda}{2}+Q_L(1+\lambda Q_R)}{(1+\lambda Q_R)^2 +\lambda^2 T_v^2}dUdy-\nonumber\\
&-2\ell\ell_4\frac{Q_R}{(1+\lambda Q_R)^2+\lambda^2 T_v^2}dVdy+  \ell^2_4\sum_{i=8}^{10} dy_i^2\\
B&=\frac{\ell^2}{4}\cos\theta d\phi\wedge d\chi +\ell \ell_4 \frac{Q_R+\lambda(Q_R^2+T_v^2)}{(1+\lambda Q_R)^2+\lambda^2 T_v^2}dy\wedge dV + \ell\ell_4 \frac{\frac{\lambda r}{2}+Q_L(1+\lambda Q_R)}{(1+\lambda Q_R)^2+\lambda^2 T_v^2}dy\wedge dU+\nonumber\\
&+\ell^2 \frac{\frac{r}{2}-Q_L Q_R +\lambda Q_L(Q_R^2+T_v^2)}{(1+\lambda Q_R)^2+\lambda^2 T_v^2}dV\wedge dU
\end{align}
\begin{align}
e^{2\Phi}=\frac{k}{p}k_4^2 \frac{1+\lambda Q_R}{(1+\lambda Q_R)^2+\lambda^2 T_v^2}\hspace{8cm}
\end{align}
One can reduce it to three dimensions, as explained in \cite{Apolo:2021wcn}, using 
\begin{align}\label{reductionansatz}
ds^2_{10}&=ds_3^2+\ell_4^2 e^{-\omega}(dy+A^{(1)})^2+d\Omega_3^2+\ell_4^2\sum_{i=8}^{10} dy_i^2\\
B_{10}&=B_3+\ell_4^2 A^{(2)}\wedge dy + \frac{\ell}{4}\cos\theta d\phi\wedge d\chi,\hspace{1cm}e^{2\Phi_{10}}=k_4^2 e^{2\Phi_3}
\end{align}
In particular, the three-dimensional reduction contains two $U(1)$ gauge fields, which resemble the two $U(1)$ gauge fields from the holographic analysis of \cite{Georgescu:2024ppd}.

\subsection{Global charges}
The backgrounds \eqref{backgrwarp} are characterized by non-trivial conserved charges that correspond to shifts in $U,V$, yielding the left and right-moving energies, shifts in $y$ and constant gauge transformations of the B-field generated by $\Lambda=dy$. After reducing to three dimensions, the last two map to gauge transformations of the two $U(1)$ gauge fields of the three-dimensional background.

These conserved charges can be computed, for example, in the covariant phase space formalism, as explained in the appendix \ref{appA:covphsp}. The result is:
\begin{align}\label{globalcha}
E_L&:=Q_{\frac{1}{\ell}\partial_U}=\frac{c}{6\ell}(T_u^2+Q_L^2),\hspace{1cm}E_R:=Q_{-\frac{1}{\ell}\partial_V}=\frac{c}{6\ell}\frac{(T_v^2+Q_R^2)(1-\lambda Q_L)}{1+\lambda Q_R}\nonumber\\
Q_{\partial_y}&=\frac{c\ell_4}{6\ell}\frac{Q_L+Q_R}{1+\lambda Q_R},\hspace{2.4cm}Q_{dy}=\frac{c\ell_4}{6\ell}\frac{Q_R-Q_L-2\lambda Q_L Q_R}{1+\lambda Q_R}+\lambda \ell_4 E_R
\end{align}
where we traded the three-dimensional Newton's constant for the Brown-Henneaux central charge $G_3=\frac{3\ell}{2c}$ \cite{Brown:1986nw}. As usually, the conserved charges are determined up to integration constants in phase space. By setting $\lambda=0$, we recover the results for the charged BTZ background, which partially motivates our choice of such integration constants. The conserved charges $Q_{\partial_y}$ and $Q_{dy}$ are quantized because $y$ is compact. In order to facilitate the comparison with $J\bar{T}$-deformed CFTs, it is useful to define the following combinations:
\begin{align}\label{undefchargesbulk}
\tilde{Q}_L:=\frac{Q_{\partial_y}-Q_{dy}}{2}=\frac{c\ell_4}{6\ell}Q_L -\frac{\lambda\ell_4 E_R}{2},\hspace{0.7cm}\tilde{Q}_R:=\frac{Q_{\partial_y}+Q_{dy}}{2}=\frac{c\ell_4}{6\ell}\frac{Q_R(1-\lambda Q_L)}{1+\lambda Q_R} +\frac{\lambda\ell_4 E_R}{2}
\end{align}
and further introduce:
\begin{align}\label{chiralcharges}
\hat{Q}_L&=\frac{c\ell_4}{6\ell}Q_L,\hspace{1cm}\hat{Q}_R=\frac{c\ell_4}{6\ell}\frac{Q_R(1-\lambda Q_L)}{1+\lambda Q_R}
\end{align}
Since $\hat{Q}_{L,R}$ are obtained by removing $E_R$-dependent part of $\tilde{Q}_{L,R}$, they are not quantized. As we shall see in section \ref{section3:asysym}, they are the natural combinations that appear in the asymptotic symmetries analysis as zero modes of the left and right-moving affine conserved charges, respectively.

\subsection{Relation to $J\bar{T}$-deformed CFTs}
The Bekenstein-Hawking entropy of the \eqref{backgrwarp} backgrounds is given by:
\begin{align}
S&=2\pi \frac{c}{6}\bigg( T_u+ \frac{T_v(1-\lambda Q_L)}{1+\lambda Q_R} \bigg)
\end{align}
We can express the parameters of the backgrounds in terms of the left and right-moving energies and the conserved charges $\hat{Q}_{L,R}$:
\begin{align}
T_u^2&=\frac{6\ell}{c}\bigg(E_L-\frac{6\ell}{c\ell_4^2}\hat{Q}^2_L\bigg),\hspace{1cm}T_v^2=\frac{6\ell}{c}\frac{E_R}{1-\lambda\frac{6\ell}{c\ell_4}(\hat{Q}_L+\hat{Q}_R)}-\frac{(\frac{6\ell}{c\ell_4})^2 \hat{Q}_R^2}{(1-\lambda\frac{6\ell}{c\ell_4}(\hat{Q}_L+\hat{Q}_R))^2}\nonumber\\
Q_L&=\frac{6\ell}{c\ell_4}\hat{Q}_L,\hspace{2.7cm} Q_R=\frac{\frac{6\ell}{c\ell_4} \hat{Q}_R}{1-\lambda\frac{6\ell}{c\ell_4}(\hat{Q}_L+\hat{Q}_R)}
\end{align}
Plugging them into the entropy formula, we obtain:
\begin{align}\label{entropynsns}
S&=2\pi \bigg[\sqrt{\frac{c}{6}\ell E_L - \frac{\ell^2}{\ell_4^2}\hat{Q}_L^2}+\sqrt{\ell E_R\bigg(\frac{c}{6}-\lambda\frac{\ell}{\ell_4}(\hat{Q}_R+\hat{Q}_L)\bigg)-\frac{\ell^2}{\ell_4^2}\hat{Q}_R^2}\bigg]
\end{align}
for which we remind that $c=6 kp$. We compare the expression with the entropy formula for a symmetric product orbifold of $J\bar{T}$-deformed 2d CFT with $p$ copies (``single-trace $J\bar{T}$"), with $c=6k$ the central charge of the seed, $k_{KM}$ the level of the $U(1)$ Kac-Moody algebra of the seed and $\lambda_{J\bar{T}}$ the deformation parameter for the seed:
\begin{align}
S_{J\bar{T}}&=2\pi\bigg[\sqrt{\ell k p E_L-\frac{k q_L^2}{k_{KM}}}+\sqrt{\ell E_R\bigg(k p-\lambda_{J\bar{T}} \frac{k (q_L-q_R)}{ \ell}\bigg)-\frac{k q_R^2}{k_{KM}}}\bigg]
\end{align}
The two expressions perfectly match if we identify $k_{KM}=k_4$, $q_L=\hat{Q}_L,q_R=-\hat{Q}_R$ and $\lambda_{J\bar{T}}=\lambda \ell_s^2/\ell_4$ \footnote{The TsT parameter $\lambda$ is dimensionless, while the $J\bar{T}$ deformation parameter has dimension of length.}. We note that using these identifications, the combinations \eqref{undefchargesbulk} that we defined map to the $U(1)$ charges in the undeformed  symmetric product orbifold CFT, which are also quantized.

Let us emphasize that type IIB string theory on the warped BTZ background is not holographically dual to a symmetric product orbifold of $J\bar{T}$-deformed CFTs, and that the long string sector does not dominate the entropy for large $k$. Hence,  the entropy matching is a surprising observation, whose interpretation is nevertheless unclear. In the main part of this article, we inspect further the relation between \eqref{backgrwarp} and the $J\bar{T}$ deformation, by analyzing the asymptotic symmetries of the warped BTZ backgrounds and comparing them with the symmetries of single-trace $J\bar{T}$.

\section{The asymptotic symmetries}
\label{section3:asysym}
In this section, we analyze the asymptotic symmetries of the backgrounds \eqref{backgrwarp}, using the covariant phase space formalism. The starting point of the analysis is the choice of boundary conditions which define a phase space of solutions of the equations of motion, that should naturally include the backgrounds \eqref{backgrwarp}. One consistency requirement is that all the allowed transformations in the phase space correspond to finite conserved charges. This condition can be translated into constraints on the asymptotic behavior of the symplectic form in phase space, evaluated on different perturbations generated by the allowed transformations. Once the boundary conditions are imposed such that these constraints are satisfied, one can proceed to compute the symmetry algebra spanned by the conserved charges. As explained in \cite{Georgescu:2022iyx}, the computation of the algebra can be performed perturbatively around the backgrounds of interest, which is a useful simplification when a closed form expression for the whole phase space is not available.

\subsection{Allowed transformations on the constant backgrounds}

In the following, we require naturally that the phase space contains all the backgrounds \eqref{backgrwarp}, namely  that the variations of the constant parameters that characterize them, $\delta T_{u,v},\delta Q_{L,R}$ are allowed. Hence, any other allowed transformations in phase space should have, in the asymptotic limit $r\rightarrow\infty$, vanishing symplectic product with them. We consider transformations that correspond to combinations of specific diffeomorphisms and gauge transformations of the B-field and evaluate the corresponding symplectic products. The constraints that we obtain fix the form of the allowed transformations in a consistent phase space, thus the boundary conditions.

The transformations that we consider are the following:
\begin{align}\label{trstart}
\xi&=F_U(U,V)\partial_U+F_V(U,V)\partial_V+F_y(U,V)\partial_y,\hspace{1cm}\Lambda=\Lambda_y(U,V)dy
\end{align}
While this is a choice, we would like to comment on the possibility that it is too restrictive. In principle, one should first fix a gauge, which would determine the radial dependence of the transformations and eliminate the trivial transformations in phase space that would lead to vanishing charges. A natural non-trivial choice would be to impose that $(\mathcal{L}_{\xi}g)_{ra}=0$ and $(\mathcal{L}_{\xi}B+\Lambda)_{ra}=0$. The corresponding transformations, whose radial dependence is not very illuminating, would contain, in principle, an additional function $F_r(U,V)$. Our consistency requirements rule it out, since it would lead to logarithmic divergences in the symplectic form. Nevertheless, it would be interesting to investigate if there exists another choice of gauge or covariant counterterms which would allow for a non-trivial radial function. Our intuition, based on the similarities with the analysis of \cite{Georgescu:2024ppd}, is that such function should be allowed and that it would lead to central charge terms in the asymptotic symmetry algebra. It would be also interesting to explore this possibility using, for example, the method put forward in \cite{Du:2024tlu} and applied to a similar set-up in \cite{Du:2024bqk}.

In order to construct perturbatively a consistent phase space around the backgrounds \eqref{backgrwarp}, we require that the symplectic form evaluated on any two allowed perturbations $\delta_1,\delta_2$ of the fields, generically denoted by $\Phi$, satisfies
\begin{align}
\omega_{a_1...a_9}(\Phi,\delta_1\Phi,\delta_2\Phi)=o(r^0)
\end{align}
where $a_i$ are indices tangent to the  boundary and the notation $o(r^{-c})$ means that the corresponding quantities, when multiplied by $r^c$ vanishes in the limit $r\rightarrow\infty$. This condition ensures the conservation of the 
symplectic flux and is a necessary condition in order for the charges associated to the allowed transformations to be conserved. In general, one usually requires also
\begin{align}
\omega_{r a_1...a_8}(\Phi,\delta_1\Phi,\delta_2\Phi)=o(r^{-1})
\end{align}
which leads to the normalizability of the symplectic form at infinity and, in some cases, implies that the associated conserved charges are finite. As we shall see, in our case the conserved charges are always finite.

The explicit expression for the symplectic form in the NS-NS sector of type IIB supergravity is listed in the appendix \ref{appA:covphsp}. We evaluated it on  variations generated by the transformations \eqref{trstart} and $\delta=\delta T_u\partial_{T_u}+\delta T_v\partial_{T_v}+\delta Q_L\partial_{Q_L}+\delta Q_R\partial_{Q_R}$ and we obtained the following equations which constrain the allowed transformations:
\begin{align}\label{equationsorder0}
&\text{\textbullet} \hspace{0.2cm}\delta T_u :\hspace{0.5cm}\partial_V F_U^{(0)}(U,V)=0\\
&\text{\textbullet} \hspace{0.2cm}\delta T_v :\hspace{0.5cm}(\partial_U+\lambda Q_L\partial_V)\big(\ell \ell_4 F_V^{(0)}(U,V) - \lambda \Lambda_y^{(0)}(U,V)\big)=0\\
&\text{\textbullet} \hspace{0.2cm}\delta Q_L :\hspace{0.5cm}\partial_V\bigg(-\ell_4^2 F_y^{(0)}(U,V)-\ell\ell_4\lambda (Q_R^2+T_v^2)F_V^{(0)}(U,V)+(1+2\lambda Q_R +\nonumber\\
&\hspace{2cm}+\lambda^2 Q_R^2+\lambda^2 T_v^2)\Lambda_y^{(0)}(U,V)\bigg)-2\ell\ell_4 Q_L(1+\lambda Q_R)\partial_V F_U^{(0)}(U,V)=0\\
&\text{\textbullet} \hspace{0.2cm}\delta Q_R :\hspace{0.5cm}\ell\ell_4(-2Q_R -\lambda Q_R^2+\lambda T_v^2)(\partial_U+\lambda Q_L\partial_V)F_V^{(0)}(U,V)+\ell_4^2(\partial_U+\lambda Q_L\partial_V)F_y^{(0)}(U,V)+\nonumber\\
&\hspace{2cm}+(1+2\lambda Q_R +\lambda^2 Q_R^2-\lambda^2 T_v^2)(\partial_U+\lambda Q_L\partial_V)\Lambda_y^{(0)}(U,V)=0
\end{align}
We added an index ${(0)}$ to emphasize that these are the functions that parametrize the allowed transformations on the backgrounds with constant parameters. The solution to these equations, up to integration constants, is given by:
\begin{align}\label{solutiontrct}
F_U^{(0)}(U,V)&=f(U),\hspace{2cm}F_V^{(0)}(U,V)=\bar{f}(v)-\frac{\lambda}{\ell \ell_4}(\eta(U)+\bar{\eta}(v))\\
F_y^{(0)}(U,V)&=-\frac{\lambda \ell(T_v^2+Q_R^2)}{\ell_4}\bar{f}(v)+\frac{\eta(U)}{\ell_4^2}-\frac{(1+2\lambda Q_R)}{\ell_4^2}\bar{\eta}(v),\hspace{0.35cm}\Lambda_y^{(0)}(U,V)=-\eta(U)-\bar{\eta}(v)
\end{align}
where we denoted by $v$ the following parameter-dependent coordinate:
\begin{align}\label{fielddepcoo0}
v&=\frac{V-\lambda Q_L U}{1+\lambda Q_R}
\end{align}
whose normalization, not fixed by the equations, was chosen for later convenience. The coordinate is identified as:
\begin{align}
v\sim v+2\pi R_v,\hspace{0.8cm}R_v=\frac{1-\lambda Q_L}{1+\lambda Q_R}
\end{align}
In order to preserve the periodicities of the spacetime coordinates $\{U,V,y\}$ and the quantized electric charge, all the functions $f(U),\bar{f}(v),\eta(U),\bar{\eta}(v)$ need to be periodic in their arguments. We note, in particular, that right-moving translations are reproduced by certain combinations of $\bar{f},\eta,\bar{\eta}$.

One can check that the allowed transformations reproduce the symmetry transformations found in the holographic analysis of $J\bar{T}$ deformed CFTs of \cite{Georgescu:2024ppd}, for backgrounds with constant parameters, which we summarize in the appendix \ref{appB:JTbar}. In particular, $v$ is the field-dependent coordinate from \cite{Georgescu:2024ppd}, which does not contain any zero mode. The only aspect of \cite{Georgescu:2024ppd} which is not reproduced by these equations solely is the absence of certain zero modes of the functions in $F_y^{(0)}$. As mentioned, at this point of the analysis we have the freedom to change $F_y^{(0)}$, as well as $F_{U,V}^{(0)}$, $\Lambda_y^{(0)}$, by any constant terms. In the following, we compute the conserved charges associated to the allowed transformations, which  indicate a particular choice of constant of $F_y^{(0)}$, based on integrability in phase space, which matches the $J\bar{T}$ analysis.

\subsection{Conserved charges}
\label{section32:cons}
\subsubsection*{Conserved charges on constant backgrounds}

We would like now to compute the conserved charges associated to the allowed transformations \eqref{solutiontrct}, using the covariant phase space formalism. As summarized in the appendix \ref{appA:covphsp}, the charge difference between two backgrounds that differ by $\delta T_{u,v},\delta Q_{L,R}$ receives contributions from the metric and the B-field, which sum up to give: 
\begin{align}\label{resultzeroordch}
\slash{\!\!\!\delta}  Q_{f,\bar{f},\eta,\bar{\eta}}=\int_0^{2\pi} d\sigma \bigg( &\frac{\ell f(U)\delta(T_u^2+Q_R^2)}{8\pi G_3} - R_v\frac{\ell \bar{f}(v)\delta(T_v^2+Q_R^2)}{8\pi G_3}+
\nonumber\\
&+\frac{\eta(U)\delta Q_L}{4\pi G_3 \ell_4} -R_v\frac{\bar{\eta}(v)\delta Q_R}{4\pi G_3 \ell_4} \bigg) 
\end{align}
It is clear that only the zero modes of the functions will contribute to the final result. Naively, the expression above does not seem to be integrable because of the factors of $R_v$ from the coefficients of $\bar{f}(v),\bar{\eta}(v)$. However, this is only an apparent issue. We can consider instead combinations of the transformations, which would correspond to a change of basis for the generators, such that the associated charge differences are integrable. For example, the following combination of constant transformations:
\begin{align}
[\bar{\eta}]_{zm}=-[\eta]_{zm}=-\frac{\lambda \ell\ell_4(T_v^2+Q_R^2)}{2(1+\lambda Q_R)}[\bar{f}]_{zm}
\end{align}
leads to an integrable expression, proportional to the right-moving energy,  $-E_R [\bar{f}]_{zm}$. More generally, one can choose combinations of the zero modes of the functions such that the transformations considered are proportional to the isometries generated by $\partial_U,\partial_V,\partial_y$ and B-field gauge transformations generated by $dy$, which should reproduce \eqref{globalcha}, up to proportionality constants. Afterwards, one can reconsider combinations of the already integrated charges to any other basis. We thus conclude that in our case the apparent issue of non-integrability is not physical and can be cured by working in a convenient basis when computing the conserved charges.\footnote{Such apparent issues regarding non-integrability appear whenever the allowed transformations in phase space are field-dependent. As explained in \cite{Compere:2015knw}, in this case, the notion of integrability should be changed accordingly, such that overall the field-dependence is not taken into account.}

It is nevertheless interesting to ask whether there is something special regarding those bases in which the charges are integrable in the first place.\footnote{The reason we ask this question is because in the similar analysis of \cite{Georgescu:2022iyx}, the conserved charges in $T\bar{T}$-deformed CFTs were reproduced from the bulk dual considering bases in which the bulk charge variations are automatically integrable.} Regarding this question, we note that by keeping only the zero modes of the functions, we would obtain, by linearity:
\begin{align}\label{exprchor0}
&\slash{\!\!\!\delta}  Q_{f,\bar{f},\eta,\bar{\eta}}=[f]_{zm} \ell\delta E_L - \bigg([\bar{f}]_{zm}-\frac{\lambda}{\ell\ell_4}[\eta+\bar{\eta}]_{zm}\bigg)\ell \delta E_R + \nonumber\\
&+\bigg(-\frac{\lambda\ell(T_v^2+Q_R^2)}{\ell_4}[\bar{f}]_{zm}+\frac{[\eta-\bar{\eta}]_{zm}}{\ell_4^2}-\frac{2\lambda Q_R}{\ell_4^2}[\bar{\eta}]_{zm}\bigg) \delta Q_{\partial_y} -[\eta+\bar{\eta}]_{zm} \delta Q_{dy}
\end{align}
Clearly, the non-integrability originates from the field-dependence of the transformations, namely from the field-dependent zero modes in $F_y^{(0)}$. We can then use the arbitrariness in the overall constant in $F_y^{(0)}$ to change the function to:
\begin{align}\label{modintegrab}
F_y^{(0)}(U,V)&=-\frac{\lambda \ell(T_v^2+Q_R^2)}{\ell_4}(\bar{f}(v)-\bar{f}_0)+\frac{\eta(U)-\bar{\eta}(v)}{\ell_4^2}-\frac{2\lambda Q_R}{\ell_4^2}(\bar{\eta}(v)-\bar{\eta}_0)
\end{align}
where we denoted by $\bar{f}_0,\bar{\eta}_0$ the zero modes of $\bar{f}$ and $\bar{\eta}$ respectively. With this modification, the charge variation becomes automatically integrable in the usual sense, and the allowed transformations match perfectly those of the $J\bar{T}$ analysis of \cite{Georgescu:2024ppd} from the appendix \ref{appB:JTbar}. In light of the discussion above, we view this merely as an interesting remark.

%We note that another resolution to the non-integrability of the expressions \eqref{resultzeroordch} is to rescale the zero modes of $\bar{f},\bar{\eta}$ by $R_v$. As we shall see, such rescaling does not affect the asymptotic symmetry algebra. 

\subsubsection*{Conserved charges on perturbed backgrounds}

In order to obtain conserved charges that are generically non-zero, we need to compute the charge differences for  backgrounds with parameters that are not constant. We can generate such backgrounds by perturbing \eqref{backgrwarp} with an  allowed transformation \eqref{solutiontrct} parametrized by $(\xi_b,\Lambda_b):=\{f_b(U),\bar{f}_b(v),\eta_b(U),\bar{\eta}_b(v)\}$:
\begin{align}
g_{\mu\nu}&=g^{(0)}_{\mu\nu}+\epsilon (\mathcal{L}_{\xi_b}g)_{\mu\nu},\hspace{1cm}B_{\mu\nu}=B_{\mu\nu}^{(0)}+\epsilon(\mathcal{L}_{\xi_b}B+\Lambda_b)_{\mu\nu}
\end{align}
where $\epsilon$ is a small parameter and $g_{\mu\nu}^{(0)},B_{\mu\nu}^{(0)}$ are the ones in \eqref{backgrwarp}. We only keep terms up to linear order in $\epsilon$ in the following computations. We obtain the following charge difference of the conserved charge associated to another transformation parametrized by $(\xi,\Lambda)=\{f(U),\bar{f}(v),\eta(U),\bar{\eta}(v)\}$, between the perturbed and the unperturbed backgrounds:
\begin{align}\label{conschargeor1}
\delta_{(\xi_b,\Lambda_b)}Q_{(\xi,\Lambda)}&=\frac{\epsilon}{4\pi G_3}\int_0^{2\pi} d\sigma \bigg(   \frac{\eta(U)}{\ell\ell_4^2}(\eta'_b+\ell\ell_4 Q_L f'_b)+\frac{f(U)}{\ell_4}(\ell\ell_4(T_u^2+Q_L^2)f'_b+Q_L \eta'_b)-\nonumber\\
&-R_v\frac{\bar{\eta}(v)}{\ell\ell_4^2}(\ell\ell_4 Q_R \bar{f}'_b+\bar{\eta}'_b)-R_v\frac{\bar{f}(v)}{\ell_4}(\ell\ell_4(T_v^2+Q_R^2)\bar{f}'_b+Q_R\bar{\eta}'_b)\bigg)
\end{align}
This expression takes the same form as the one of the conserved charges in the holographic dual of a $J\bar{T}$-deformed CFT, linearized around the background with constant parameters \eqref{conservedchJTbarlin}. It is convenient for later to introduce separate notations for the various terms in \eqref{conschargeor1} {\footnote{We drop  the $\delta$ from the notation for simplicity, but we should understand them as infinitesimal charge variations, since our expressions are only valid linearly around the constant backgrounds}}, $P_{\eta},\bar{P}_{\bar{\eta}},Q_f,\bar{Q}_{\bar{f}}$, each corresponding to the case when only the function appearing in its index is non-zero. We can include also  the order 0 results in order to write:
\begin{align}
P_{\eta}&=\frac{1}{4\pi G_3\ell_4}\int_0^{2\pi} d\sigma \eta(U)\bigg[ Q_L+\epsilon \bigg(\frac{\eta'_b}{\ell\ell_4}+Q_L f'_b\bigg)\bigg]\\
Q_f&=\frac{\ell}{8\pi G_3}\int_0^{2\pi} d\sigma f(U)\bigg[ T_u^2+Q_L^2 +\epsilon\bigg( 2(T_u^2+Q_L^2)f'_b + \frac{2 Q_L}{\ell\ell_4}\eta'_b\bigg) \bigg]\\
\bar{P}_{\bar{\eta}}&=-\frac{1}{4\pi G_3 \ell_4} \int_0^{2\pi} d\sigma \bar{\eta}(v)R_v \bigg[ Q_R + \epsilon\bigg(\frac{\bar{\eta}'_b}{\ell\ell_4}+Q_R \bar{f}'_b \bigg)\bigg]\\
\bar{Q}_{\bar{f}}&=-\frac{\ell}{8\pi G_3}\int_0^{2\pi} d\sigma \bar{f}(v) R_v\bigg[ T_v^2+Q_R^2 + \epsilon\bigg( 2(T_v^2+Q_R^2)\bar{f}'_b+\frac{2 Q_R}{\ell\ell_4}\bar{\eta}'_b \bigg)\bigg]
\end{align}
We will refer to $P_{\eta},\bar{P}_{\bar{\eta}}$ as left and right-moving affine charges respectively, to $Q_f$ as (left-moving) conformal charges and to $\bar{Q}_{\bar{f}}$ as (right-moving) pseudo-conformal charges, using the terminology from $J\bar{T}$-deformed CFTs, which will be motivated by the computation of their algebra.

We would like to compute perturbatively in the small parameter $\epsilon$ the asymptotic symmetry algebra of the conserved charges, on the perturbed backgrounds. As explained in detail in \cite{Georgescu:2022iyx}, such computation requires the knowledge of the allowed transformations on the perturbed backgrounds, which are of the form:
\begin{align}\label{trseriesper}
\xi&=\xi^{(0)}+\epsilon \xi^{(1)}\hspace{1cm}\Lambda=\Lambda^{(0)}+\epsilon \Lambda^{(1)}
\end{align}
with $\xi^{(0)},\Lambda^{(0)}$ the allowed transformations on the constant backgrounds that we derived previously and $\xi^{(1)},\Lambda^{(1)}$ the corrections that we have to determine. We would like to mention that a modification of the background functions $(\xi_b,\Lambda_b)$ by constant terms, like the one proposed in \eqref{modintegrab}, does not affect the equations that we obtain below, so we can simply use \eqref{solutiontrct} in the following.

\subsection{Allowed transformations on the perturbed backgrounds}
\label{section33:allowedperturbed}
In order to derive the corrections $\xi^{(1)},\Lambda^{(1)}$, we evaluate again the symplectic form between variations generated by the transformations \eqref{trseriesper} and a general variation of the constant parameters, $\delta=\delta T_u\partial_{T_u}+\delta T_v\partial_{T_v}+\delta Q_L\partial_{Q_L}+\delta Q_R\partial_{Q_R}$, but now up to linear order in $\epsilon$ on top of the perturbed backgrounds parametrized by $(\xi_b,\Lambda_b)$. We expand to linear order the functions:
\begin{align}
F_U&=F_U^{(0)}(U,V)+\epsilon F_U^{(1)}(U,V),\hspace{1cm}F_V=F_V^{(0)}(U,V)+\epsilon F_V^{(1)}(U,V)\\
F_y&=F_y^{(0)}(U,V)+\epsilon F_y^{(1)}(U,V),\hspace{1cm}\Lambda_y=\Lambda_y^{(0)}(U,V)+\epsilon \Lambda_y^{(1)}(U,V)
\end{align}
where the order 0 terms are given in \eqref{solutiontrct}. Clearly, there are two types of contributions:
\begin{itemize}
\item allowed transformations at order $\epsilon$ on the constant backgrounds
\item allowed transformations at order $0$ on the backgrounds at order $\epsilon$
\end{itemize}
The first contribution just gives the LHS of the equations \eqref{equationsorder0}, but for the functions that appear at order $\epsilon$ (i.e. denoted by $^{(1)}$), while the RHS is no longer 0 because of the second type of contributions for the perturbed backgrounds. 

We list the constraints that we obtain in \eqref{ecper1}-\eqref{ecper4} in appendix \ref{appC:details}. We present there the details of solving these equations and we give below the main result. One key observation that we use is that the corrections $F_U^{(1)},F_V^{(1)},F_y^{(1)},\Lambda_y^{(1)}$ should be determined in terms of the functions $f,\bar{f},\eta,\bar{\eta}$ and those parametrizing the background solely. The reason is that the sets of allowed transformations at different points in phase space should be in 1-to-1 relation. This remark allows us to identify the variations of the constant parameters generated by the action of the zeroth order transformations as:
\begin{align}\label{variations2}
\delta T_u^2&=2 T_u^2 f'_b,\hspace{0.5cm}\delta T_v^2=2 T_v^2 \bar{f}'_b,\hspace{0.5cm}\delta Q_L=\frac{\eta'_b}{\ell\ell_4}+Q_L f'_b,\hspace{0.5cm}\delta Q_R=\frac{\bar{\eta}'_b}{\ell\ell_4}+Q_R \bar{f}'_b
\end{align}
and the variation of the field-dependent coordinate as:
\begin{align}\label{var4}
\delta v&= -\frac{\lambda}{\ell\ell_4(1+\lambda Q_R)}\bigg(\ell\ell_4 Q_L (f_b-f_{b0}) + \ell\ell_4 Q_R (\bar{f}_b-\bar{f}_{b0})+(\eta_b-\eta_{b0})+(\bar{\eta}_b-\bar{\eta}_{b0}) + \mathcal{V}_0\bigg)
\end{align}
where $\mathcal{V}_0$ is an integration constant. The final result for the allowed  transformations on top of the perturbed backgrounds is:
\begin{align}
F_V&=\bar{f}-\frac{\lambda}{\ell\ell_4}(\eta+\bar{\eta})+\epsilon\bigg(\bar{f}'-\frac{\lambda}{\ell\ell_4}
\bar{\eta}'\bigg)\delta v-\epsilon \frac{\lambda}{\ell\ell_4}w_{\bar{f},\bar{\eta}}\bigg(U-\frac{v}{R_v}\bigg)\\
F_y&=-\frac{\lambda \ell(T_v^2+Q_R^2)}{\ell_4}\bar{f}+\frac{\eta}{\ell_4^2}-\frac{(1+2\lambda Q_R)}{\ell_4^2}\bar{\eta}+\epsilon\bigg(-\frac{\lambda \ell(T_v^2+Q_R^2)}{\ell_4}\bar{f}'-\frac{(1+2\lambda Q_R)}{\ell_4^2}\bar{\eta}'\bigg)\delta v-\nonumber\\
&-\epsilon \frac{2}{\ell\ell_4^3}\int^v \bigg(\lambda \bar{\eta}'_b(\ell\ell_4 Q_R\bar{f}'+\bar{\eta}')+\ell\ell_4\lambda \bar{f}'_b(\ell\ell_4(T_v^2+Q_R^2)\bar{f}'+Q_R\bar{\eta}')\bigg)+\nonumber\\
&+\epsilon\frac{w_{\bar{f},\bar{\eta}}}{\ell_4^2}\bigg(U+(1+2\lambda Q_R)\frac{v}{R_v}\bigg)\\
\Lambda_y&=-(\eta(U)+\bar{\eta}(v))+\epsilon\bigg[\bar{\eta}'\delta v-w_{\bar{f},\bar{\eta}}\bigg(U-\frac{v}{R_v}\bigg)\bigg]\hspace{4cm}
\end{align}
where the coefficients of the terms linear in $U,v$ are determined by:
\begin{align}
w_{\bar{f},\bar{\eta}}:=&-\frac{3\lambda \ell\ell_4}{c}\frac{(\bar{P}_{\bar{\eta}'}+\bar{Q}_{\bar{f}'})}{1+\lambda Q_R}
\end{align} 
The allowed transformations agree with those of the bulk dual of a $J\bar{T}$-deformed CFT, expanded to the same order in $\epsilon$, listed in appendix \ref{appB:JTbar}. An essential feature of our result is that  whenever the periodic functions $\bar{\eta}(v)$ and/or $\bar{f}(v)$ are non-trivial, we also need to turn on a linear $\eta(U)$, as well as a linear $\bar{\eta}(v)$. The coefficients of these linear terms are charge-dependent, which predicts that the asymptotic symmetry algebra is non-linear. Before computing the algebra, we comment on the representation theorem.

\subsection{Comments on the representation theorem}
In the perturbative analysis of \cite{Georgescu:2022iyx}, the functions at linear order were fixed using the representation theorem of \cite{Barnich:2007bf,Barnich:2010eb}. However, in order for the representation theorem to hold at the order considered, \cite{Georgescu:2022iyx} had to slightly modify the bracket of the allowed transformations by removing field-dependent zero modes. This modification was in fact required in order for the standard bracket to close, which is a necessary, but not sufficient requirement in order for the representation theorem to hold. The closure of the modified bracket seemed to be related to the requirement of integrability of the charge variations, which is used in the proof of the representation theorem, but no precise statement was formulated.

In the following, we would like to check whether the standard bracket of allowed transformations
\begin{align}
[(\xi_1,\Lambda_1),(\xi_2,\Lambda_2)]&=(\xi,\Lambda)
\end{align}
where $\xi$ is given by the modified Lie bracket of vector fields:
\begin{align}\label{modlie}
\xi&=[\xi_1,\xi_2]_{L.B.}-(\delta_1\xi_2-\delta_2\xi_1)
\end{align}
and $\Lambda$ is given by the modified bracket:
\begin{align}\label{modliegauge}
\Lambda&=\mathcal{L}_{\xi_1}\Lambda_2-\mathcal{L}_{\xi_2}\Lambda_1-(\delta_1\Lambda_2-\delta_2\Lambda_1)
\end{align}
closes, at linear order in $\epsilon$.

We first focus on the variation:
\begin{align}\label{varprim0}
\delta_1\xi_2&=\bigg(\bar{f}'_2-\frac{\lambda}{\ell\ell_4}\bar{\eta}'_2\bigg)\delta_1 v \partial_V+\bigg(-\frac{\lambda\ell}{\ell_4} (T_v^2+Q_R^2)\bar{f}'_2-\frac{1+2\lambda Q_R}{\ell_4^2}\bar{\eta}'_2\bigg)\delta_1 v \partial_y+\nonumber\\
&+\int^v \bigg(-\frac{\lambda \ell}{\ell_4}\bar{f}'_2 \delta_1(T_v^2+Q_R^2)-\frac{2\lambda \bar{\eta}'_2}{\ell_4^2}\delta_1 Q_R \bigg) 
\end{align}
After using the explicit expressions for the variations \eqref{var1}, we see that the term in the second line is symmetric in $1\leftrightarrow 2$, so it drops in the combination $\delta_1\xi_2-\delta_2\xi_1$. We note that the result differs from the one we would have obtained by naively varying:
\begin{align}
\delta_1\bigg(-\frac{\lambda\ell}{\ell_4}(T_v^2+Q_R^2)\bar{f}_2-\frac{2\lambda Q_R \bar{\eta}_2}{\ell_4^2}\bigg)=-\frac{\lambda\ell}{\ell_4}\delta_1(T_v^2+Q_R^2)\bar{f}_2-\frac{2\lambda \delta_1 Q_R \bar{\eta}_2}{\ell_4^2}
\end{align}
After computing the Lie bracket of the vector fields, we find that $\xi$ defined in \eqref{modlie} is indeed of the form \eqref{solutiontrct}, with:
\begin{align}\label{functionsmod}
f&=f_1 f'_2-f_2 f'_1, \hspace{1cm}\bar{f}=\bar{f}_1 \bar{f}'_2-\bar{f}_2 \bar{f}'_1+\frac{\lambda}{\ell\ell_4(1+\lambda Q_R)}(v_{01}\bar{f}'_2-v_{02}\bar{f}'_1)\\
\eta&=f_1\eta'_2-f_2\eta'_1, \hspace{1cm}\bar{\eta}=\bar{f}_1\bar{\eta}'_2-\bar{f}_2\bar{\eta}'_1+\frac{\lambda}{\ell\ell_4(1+\lambda Q_R)}(v_{01}\bar{\eta}'_2-v_{02}\bar{\eta}'_1)
\end{align}
where $v_{01,2}=\delta_{1,2}v_0$ and $v_0$ is defined via $\delta v= -\frac{\lambda}{\ell\ell_4(1+\lambda Q_R)}(\ell\ell_4 Q_L f + \ell\ell_4 Q_R \bar{f}+\eta+\bar{\eta}  +v_0)$. It is easy to check that also $\Lambda$ defined in \eqref{modliegauge} has the $y$ component of the form $\Lambda_y=-(\eta+\bar{\eta})$, with the two functions in \eqref{functionsmod}. The resulting gauge transformation has, nevertheless, extra components $\Lambda_{u,v}$ which change the three-dimensional B-field $B_{uv}$. We did not take into account these kind of transformations in our analysis and restricted our attention to diffeomorphisms and gauge transformations of the two three-dimensional gauge fields.

We note that the presence of the zero mode $v_0$, in fact of any constant in $\delta v$, has the effect to modify $\bar{f}_{1,2}\rightarrow \bar{f}_{1,2}+\frac{\lambda}{\ell\ell_4(1+\lambda Q_R)}v_{01,2}$. When mapping the transformations to conserved charges, such terms proportional to $v_{01,2}$ in \eqref{functionsmod} give 0, since the functions that appear are periodic functions without zero modes. Hence, such transformations are not physical, at least at this order of the analysis, and should be dropped, which amounts for removing completely the zero mode of $\delta v$, namely setting $\mathcal{V}_0=0$ in \eqref{var4}.

Considering the rest of the terms, one can easily check that the representation theorem holds at this order, by comparing the results with \eqref{orderzeroalgebra} and ignoring the commutators that are pure central charge and thus are not visible in the algebra of allowed transformations. We emphasize again that the knowledge about the primitive in \eqref{varprim0} was crucial for this check.

Finally, let us discuss the case of allowed transformations with the field-dependent zero modes of $F_y^{(0)}$ removed \eqref{modintegrab}. If we require that the primitive in \eqref{varprim0} has no zero mode, the result for $\xi$ clearly does not change, since the corresponding term is exactly 0. However, the form of the allowed transformations now changed, since we require no field-dependent zero mode in the resulting $y$ components of $\xi$. Hence, in order for the modified bracket to hold, one needs to further modify it by removing the zero modes of $\bar{f},\bar{\eta}$ in the field-dependent terms, as in \eqref{modintegrab}. This extra modification is exactly the one required in the $T\bar{T}$ analysis of \cite{Georgescu:2022iyx}. An equivalent way of formulating this modification is to require a particular variation of the zero modes, namely:
\begin{align}
\delta_1\bigg( -\frac{\lambda\ell(T_v^2+ Q_R^2)}{\ell_4}[\bar{f}_{2}]_{zm}-\frac{2\lambda Q_R}{\ell_4^2}[\bar{\eta}_{2}]_{zm} \bigg)&=\frac{\lambda\ell(Q_R^2+T_v^2)}{\ell_4}[\bar{f}_1\bar{f}'_2]_{zm}+\frac{2\lambda Q_R}{\ell_4^2}[\bar{f}_1\bar{\eta}'_2]_{zm}
\end{align}
It would be interesting to understand the interpretation of such a variation of the zero modes, which now need to be considered field-dependent.

\subsection{The charge algebra}
After we obtained the allowed transformations, we are ready to compute the asymptotic symmetry algebra of their associated conserved charges. The Poisson bracket between conserved charges is written generically as:
\begin{align} \label{definitionpoissonbr}
\{Q_{\xi,\Lambda},Q_{\chi,\Sigma}\}&\equiv \delta_{\chi,\Sigma}Q_{\xi,\Lambda}
\end{align}
where $\xi, \chi$ stand for diffeomorphisms and $\Lambda,\Sigma$ for the associated gauge transformations.

 We compute first the order 0 contribution, namely the charge algebra on constant backgrounds, in which the only non-zero Poisson brackets are those which yield a global charge. We obtain:
\begin{align} \label{orderzeroalgebra}
\{Q_{f_1},Q_{f_2}\}&=\frac{\ell (T_u^2+Q_L^2)}{4\pi G_3}\int_0^{2\pi} \hspace{-0.2cm}d\sigma\; f_1 f'_2,\hspace{0.6cm}\{\bar{Q}_{\bar{f}_1},\bar{Q}_{\bar{f}_2}\}=-R_v\frac{\ell (T_v^2+Q_R^2)}{4\pi G_3}\int_0^{2\pi} \hspace{-0.2cm}d\sigma\; \bar{f}_1 \bar{f}'_2\\
\{P_{\eta_1},P_{\eta_2}\}&=\frac{1}{4\pi G_3\ell\ell_4^2}\int_0^{2\pi} \hspace{-0.2cm}d\sigma\; \eta_1 \eta'_2,\hspace{1cm}\{\bar{P}_{\bar{\eta}_1},\bar{P}_{\bar{\eta}_2}\}=-R_v\frac{1}{4\pi G_3\ell\ell_4^2}\int_0^{2\pi} \hspace{-0.2cm}d\sigma\; \bar{\eta}_1 \bar{\eta}'_2\\
\{Q_{f_1},P_{\eta_2}\}&=\frac{Q_L}{4\pi G_3\ell_4}\int_0^{2\pi} \hspace{-0.2cm}d\sigma\; f_1 \eta'_2,\hspace{1cm}\{\bar{Q}_{\bar{f}_1},\bar{P}_{\bar{\eta}_2}\}=-R_v\frac{Q_R}{4\pi G_3\ell_4}\int_0^{2\pi} \hspace{-0.2cm}d\sigma\; \bar{f}_1 \bar{\eta}'_2
\end{align}
while all the other Poisson brackets vanish. By expanding the periodic functions in a Fourier basis, $f,\eta=e^{i m U}$, $\bar{f},\bar{\eta}=e^{-i m v/R_v}$, it is clear that the only non-zero Poisson brackets are: 
\begin{align}
\{Q_m,Q_{-m}\}&=-2 i m Q_0, \hspace{1cm}\{\bar{Q}_m,\bar{Q}_{-m}\}=\frac{2 i m}{R_v} \bar{Q}_0\\
\{P_m,P_{-m}\}&=-2 i m \frac{c}{6},\hspace{1.2cm}\{\bar{P}_m,\bar{P}_{-m}\}=-2 i m \frac{c}{6}\\
\{Q_m,P_{-m}\}&=-2 i m P_0,\hspace{1cm}\{\bar{Q}_m,\bar{P}_{-m}\}=2 i m \bar{P}_0
\end{align}
where we remind that the order 0 charges  are given by:
\begin{align}
Q_0&=\frac{c}{6}(T_u^2+Q_L^2),\hspace{0.7cm}\bar{Q}_0=-\frac{c}{6}R_v(T_v^2+Q_R^2),\hspace{0.7cm}P_0=\frac{c}{6}\frac{Q_L}{\ell\ell_4},\hspace{0.7cm}\bar{P}_0=-\frac{c}{6}\frac{Q_R R_v}{\ell\ell_4}
\end{align}
Here $Q_0,\bar{Q}_0$ are dimensionless versions of \eqref{globalcha} and $P_0=\frac{\hat{Q}_L}{\ell_4^2},\bar{P}_0=-\frac{\hat{Q}_R}{\ell_4^2}$ are rescaled versions of \eqref{chiralcharges}. We note that these Poisson brackets are consistent with the symmetry algebra of $J\bar{T}$ deformed CFTs. In order to obtain all the non-trivial Poisson brackets we need to perform the computations on top of the perturbed backgrounds, where we need to take in account the additional contribution to the asymptotic Killing vectors and their conserved charges.

This extra terms in the algebra are obtained by plugging in \eqref{definitionpoissonbr} the allowed transformations on perturbed backgrounds. Although we will not write this explicitly, all the Poisson brackets computed are valid be up to linear order in the small parameter $\epsilon$ introduced previously.
It is useful to compute the Poisson brackets separately for the various independent periodic functions that we can turn on. We compare the results that we obtain to the symmetry algebra of a symmetric product orbifold of $J\bar{T}$-deformed CFTs \cite{Chakraborty:2023wel}.

\subsubsection*{Left affine bracket $\{P_{\eta_1},P_{\eta_2}\}$}
The Poisson bracket between two left-moving affine charges does not receive any correction at linear order. Hence, it only consists of a central charge term. We copy the result here, after expanding the functions in the Fourier basis:
\begin{align}
\{P_m,P_n\}&=-\frac{i m c}{3 \ell^2\ell^2_4} \delta_{m+n,0}
\end{align}
In order to work with dimensionless quantities, let us rescale the charges  $P_m\rightarrow \frac{\ell^2_4 P_m}{2}$ such that:
\begin{align}
\{P_m,P_n\}&=-\frac{i m p k_4}{2} \delta_{m+n,0}
\end{align}
where we used that $c=6 k p$. The choice of rescaling factor $\ell_4^2$ is motivated by the relation between the global charges $P_0,\bar{P}_0$ and $\hat{Q}_{L,R}$.

The corresponding Poisson bracket in the symmetry algebra of a $J\bar{T}$-deformed CFT is:
\begin{align}
\{P_m^{J\bar{T}},P_n^{J\bar{T}}\}&=-\frac{i m k_{KM}}{2}\delta_{m+n,0}
\end{align}
A symmetric product orbifold of $J\bar{T}$-deformed CFTs with $p$ copies and $U(1)$ Kac-Moody level of the seed theory $k_{KM}$ also had a $U(1)$ left-moving Kac-Moody symmetry algebra whose level is $p k_{KM}$ \cite{Chakraborty:2023wel}. Hence, by identifying $k_4=k_{KM}$ as instructed by the entropy matching reviewed in section \ref{section2:review}, we match this Poisson bracket computed for the warped BTZ backgrounds to a bracket in the symmetry algebra of single-trace $J\bar{T}$.

\subsubsection*{Left affine - right affine  bracket $\{P_{\eta_1},\bar{P}_{\bar{\eta}_2}\}$}
The Poisson bracket between a left and a right-moving affine charge can be written as:
\begin{align}
\{P_{\eta_1},\bar{P}_{\bar{\eta}_2}\}=\delta_{\bar{\eta}_2} P_{\eta_1}&=\delta_{\bar{\eta}_2(p)} P_{\eta_1}+\delta_{\bar{\eta}_2(w)} P_{\eta_1}
\end{align}
where we separated the terms coming from the periodic function $\bar{\eta}_2$ that appears in the allowed transformations, labeled by $(p)$ and the terms linear in $U,v$, proportional to the winding, labeled by $(w)$. We obtain:
\begin{align}
\delta_{\bar{\eta}_2(p)}P_{\eta_1}&=0\\
\delta_{\bar{\eta}_2(w)}P_{\eta_1}&=\int_0^{2\pi}d\sigma \frac{w_2 \eta_1}{4\pi G_3\ell\ell_4^2}
\end{align}
where we denoted for simplicity the winding by $w_2=w_{\bar{\eta}_2}$. We can write the result as:
\begin{align} \label{leftrightaffine}
\{P_{\eta_1},\bar{P}_{\bar{\eta}_2}\}=\frac{c}{3 } \frac{w_2 [\eta_1]_{zm} }{\ell^2\ell_4^2}
\end{align}
It is instructive to check that the Poisson bracket is antisymmetric, namely that we can compute the same quantity as:
\begin{align}
\{P_{\eta_1},\bar{P}_{\bar{\eta}_2}\}=-\delta_{\eta_1} \bar{P}_{\bar{\eta}_2}&=-(\delta_{\eta_1} \bar{P}_{\bar{\eta}_2(p)}+\delta_{\eta_1}  \bar{P}_{\bar{\eta}_2(w)})
\end{align}
with the two terms given by:
\begin{align}
\delta_{\eta_1} \bar{P}_{\bar{\eta}_2(p)}&=\frac{1}{4\pi G_3 \ell_4^3\ell^2}\int_0^{2\pi}\hspace{-0.3cm}d\sigma\bigg(-\ell\ell_4 P_{r2}(v) \eta'_1-\frac{\lambda R_v}{(1+\lambda Q_R)}\eta_1\bar{\eta}'_2\big(\ell\ell_4 Q_R \bar{f}'_b+\bar{\eta}'_b\big)\bigg)\\
\delta_{\eta_1} \bar{P}_{\bar{\eta}_2(w)}&=\int_0^{2\pi}\hspace{-0.3cm}d\sigma \frac{w_1 U \eta'_1}{4\pi G_3 \ell\ell_4^2}
\end{align}
We remind that, when we have only $\bar{\eta}_2$ turned on \footnote{See appendix \ref{appC:details} for the general definition of the function.}:
\begin{align}
P_{r2}(v)&=\int^v \frac{\lambda}{\ell\ell_4(1+\lambda Q_R)}\big( \ell\ell_4 Q_R \bar{f}'_b+\bar{\eta}'_b \big)\bar{\eta}'_2
\end{align}
Separating the periodic part of it from the nonperiodic part and integrating by parts, we find:
\begin{align}
\delta_{\eta_1} \bar{P}_{\bar{\eta}_2(p)}&=-\frac{1}{4\pi G_3 \ell\ell_4^2}\int_0^{2\pi}\hspace{-0.3cm}d\sigma  \;w_2(\eta_1 +\hat{v}\eta'_1)
\end{align}
where $\hat{v}=v/R_v$. Hence, we obtain:
\begin{align}
\{P_{\eta_1},\bar{P}_{\bar{\eta}_2}\}&= \frac{1}{4\pi G_3 \ell\ell_4^2}\int_0^{2\pi}\hspace{-0.3cm}d\sigma  \; w_2 (\eta_1-(u-\hat{v})\eta'_1)
\end{align}
which gives precisely \eqref{leftrightaffine}, as it should assuming antisymmetry. Finally, let us expand in Fourier modes the functions $\eta_1=e^{i m U}$ and $\bar{\eta}_2=e^{- i n v/R_v}$:
\begin{align}
\{P_m,\bar{P}_n\}&=\frac{\lambda  i n}{\ell_4 \ell (1-\lambda Q_L)} \bar{P}_n\delta_{m,0}
\end{align}
We rescale like before $P_m\rightarrow \frac{\ell_4^2 P_m}{2}$ and $P_m\rightarrow \frac{\ell_4^2 \bar{P}_n}{2}$ and finally we replace $\lambda=\lambda_{J\bar{T}}\frac{\ell_4}{\ell_s^2}$ as instructed by the entropy matching and we obtain:
\begin{align}
\{P_m,\bar{P}_n\}&=\frac{\lambda_{J\bar{T}} i n k_4 p}{2 R_Q}\bar{P}_n \delta_{m,0}
\end{align}
where we denoted $R_Q=\ell(1-\lambda Q_L)p$.

\subsubsection*{Right affine bracket $\{\bar{P}_{\bar{\eta}_1},\bar{P}_{\bar{\eta}_2}\}$}
Similarly, the Poisson bracket between two right-moving affine charges can be written as:
\begin{align}
\{\bar{P}_{\bar{\eta}_1},\bar{P}_{\bar{\eta}_2}\}=\delta_{\bar{\eta}_2} \bar{P}_{\bar{\eta}_1}&=\delta_{\bar{\eta}_2(p)} \bar{P}_{\bar{\eta}_1(p)}+\delta_{\bar{\eta}_2(p)} \bar{P}_{\bar{\eta}_1(w)}+\delta_{\bar{\eta}_2(w)} \bar{P}_{\bar{\eta}_1(p)}+\delta_{\bar{\eta}_2(w)} \bar{P}_{\bar{\eta}_1(w)}
\end{align}
where we separated again the terms coming from the periodic functions that appear in the allowed transformations, labeled by $(p)$ and the terms linear in $U,v$, proportional to the winding, labeled by $(w)$. Since the winding terms come each with a factor of $\epsilon$, we can discard the last term. The rest of the contributions proportional to $\epsilon$ yield:
\begin{align}
\delta_{\bar{\eta}_2(p)} \bar{P}_{\bar{\eta}_1(p)}&=\frac{R_v}{4\pi G_3\ell\ell_4^2}\int_0^{2\pi} d\sigma \bigg[\frac{\lambda\bar{\eta}_1\bar{\eta}'_2}{\ell\ell_4(1+\lambda Q_R)}(\ell\ell_4 Q_R \bar{f}'_b+\bar{\eta}'_b)+\bar{\eta}'_1(w_2 \hat{v} + P_{r2(p)}) \bigg]  \\
\delta_{\bar{\eta}_2(p)} \bar{P}_{\bar{\eta}_1(w)}&=-\int_0^{2\pi} d\sigma \frac{w_1 \bar{\eta}_2}{4\pi G_3 \ell\ell_4^2}\\
\delta_{\bar{\eta}_2(w)} \bar{P}_{\bar{\eta}_1(p)}&=-\int_0^{2\pi} d\sigma \frac{w_2 \hat{v} R_v \bar{\eta}'_1}{4\pi G_3\ell\ell_4^2}
\end{align}
where we denoted for simplicity the windings $w_{1,2}=w_{\bar{\eta}_{1,2}}$. We also used the notation $\hat{v}=v/R_v$ and we separated the periodic part of the primitives $P_{r1,2}(v)$ from the non-periodic one originating from the zero modes of the functions inside. We note that when we sum them up, the terms proportional to $\hat{v}$ cancel. Moreover, we can integrate by parts and use the expression for $P_{r2}(v)$ in order to obtain:
\begin{align}
\delta_{\bar{\eta}_2}\bar{P}_{\bar{\eta}_1}&=-\frac{1}{4\pi G_3\ell\ell_4^2} \int_0^{2\pi} d\sigma (w_1\bar{\eta}_2-w_2 \bar{\eta}_1)=\frac{c}{3 \ell^2\ell_4^2}(w_2[\bar{\eta}_1]_{zm}-w_1[\bar{\eta}_2]_{zm})
\end{align}
The order 0 term, which is a central charge term, does not receive any correction, except that for now it is non-zero for arbitrary modes. Thus, the final result is:
\begin{align}
\{\bar{P}_{\bar{\eta}_1},\bar{P}_{\bar{\eta}_2}\}&=-\frac{i m c}{3 \ell^2\ell^2_4} \delta_{m+n,0}+\frac{c}{3 \ell^2\ell_4^2}(w_2[\bar{\eta}_1]_{zm}-w_1[\bar{\eta}_2]_{zm})
\end{align}
We expand it into Fourier modes to write, after the same rescalings and redefinitions that we used for the previous Poisson brackets:
\begin{align}
\{\bar{P}_m,\bar{P}_n\}&=-\frac{i m p k_4}{2} \delta_{m+n,0}+ \frac{\lambda_{J\bar{T}} i n k_4 p}{2 R_Q}\bar{P}_n \delta_{m,0}-\frac{\lambda_{J\bar{T}} i m k_4 p}{2 R_Q}\bar{P}_m \delta_{n,0}
\end{align}
which matches perfectly the corresponding bracket in a symmetric product orbifold of $J\bar{T}$-deformed CFTs.

\subsubsection*{Left conformal bracket $\{Q_{f_1},Q_{f_2}\}$}
In the case of two left-moving conformal transformations there are no winding terms involved and we simply obtain at order $\epsilon$:
\begin{align}
\{Q_{f_1},Q_{f_2}\}&=-\frac{1}{4\pi G_3 \ell_4}\int_0^{2\pi}d\sigma \; (f_2 f'_1-f_1 f'_2)\big( \ell\ell_4 (T_u^2+ Q_L^2)f'_b+ Q_L \eta'_b  \big)
\end{align}
We can add also the result at order 0 and recognize that overall:
\begin{align}
\{Q_{f_1},Q_{f_2}\}&=Q_{f_1 f'_2-f_2 f'_1}
\end{align}
Expanding into Fourier modes, we obtain:
\begin{align}
\{Q_m,Q_n\}&=-i(m-n)Q_{m+n}
\end{align}
We note that we do not see the central extension that should be present because we set the radial function of the diffeomorphisms to 0, similarly to the analysis of \cite{Georgescu:2022iyx}.

\subsubsection*{Left conformal-left affine bracket $\{Q_{f_1},P_{\eta_2}\}$}
Another simple Poisson bracket is between a left-moving conformal charge and a left-moving affine one, since there is no winding involved. We obtain:
\begin{align}
\{Q_{f_1},P_{\eta_2}\}&=\delta_{\eta_2}Q_{f_1}=\frac{1}{4\pi G_3 \ell \ell^2_4}\int_0^{2\pi}d\sigma \; f_1 \eta'_2 \big( \ell\ell_4 Q_L f'_b+  \eta'_b  \big)= P_{f_1 \eta'_2}
\end{align}
which leads to the following bracket for the Fourier modes:
\begin{align}
\{Q_m,P_n\}&=i n P_{m+n}
\end{align}

\subsubsection*{Left conformal-right affine bracket $\{\bar{P}_{\bar{\eta}_1},Q_{f_2}\}$}
For a right-moving affine charge and a left-moving conformal charge, the Poisson bracket can be split into:
\begin{align}
\{\bar{P}_{\bar{\eta}_1},Q_{f_2}\}&=\delta_{f_2}\bar{P}_{\bar{\eta}_1}=\delta_{f_2}\bar{P}_{\bar{\eta}_{1(p)}}+\delta_{f_2}\bar{P}_{\bar{\eta}_{1(w)}}
\end{align}
The first contribution is given by:
\begin{align}
\delta_{f_2}\bar{P}_{\bar{\eta}_{1(p)}}&=-\frac{Q_L}{4\pi G_3 \ell\ell_4^2}\int_0^{2\pi}d\sigma\; \bigg( \ell\ell_4 P_{r1}(v) +\frac{\lambda R_v}{1+\lambda Q_R} f_2 \bar{\eta}_1 \big(\ell\ell_4 Q_R \bar{f}'_b+\bar{\eta}'_b \big) \bigg)
\end{align}
We split again the primitive into periodic and non-periodic part, we integrate by parts the periodic one and, after some cancellations, we are left with:
\begin{align}
\delta_{f_2}\bar{P}_{\bar{\eta}_{1(p)}}&=-\frac{Q_L}{4\pi G_3 \ell_4}\int_0^{2\pi}d\sigma\; w_1 \big( f_2 + \hat{v} f'_2 \big)
\end{align}
The winding contribution is given by:
\begin{align}
\delta_{f_2}\bar{P}_{\bar{\eta}_{1(w)}}&=\frac{Q_L}{4\pi G_3 \ell_4}\int_0^{2\pi}d\sigma\; w_1 U f'_2
\end{align}
The $(U-\hat{v})$ term drops in the sum and we obtain:
\begin{align}
\{\bar{P}_{\bar{\eta}_1},Q_{f_2}\}&=-\frac{c Q_L}{3 \ell_4\ell} w_1 [f_2]_{zm}
\end{align}
After expanding the functions in Fourier modes and rescaling the order 0 charges as explained, the result is:
\begin{align}
\{\bar{P}_{m},Q_{n}\}&=-\frac{i m \lambda_{J\bar{T}}}{R_Q}\bar{P}_m P_0\delta_{n,0}
\end{align}
Clearly, we only see the order 0 charge $P_0 \delta_{n,0}$ in the non-linear result at this order in the perturbative analysis.

\subsubsection*{Right pseudoconformal-left affine bracket $\{\bar{Q}_{\bar{f}_1},P_{\eta_2}\}$}
Next, we compute the Poisson bracket between a right-moving pseudoconformal charge and a left-moving affine charge:
\begin{align}
\{\bar{Q}_{\bar{f}_1},P_{\eta_2}\}&=\delta_{\eta_2} \bar{Q}_{\bar{f}_1}= -\frac{1}{4\pi G_3 \ell\ell_4^2}\int_0^{2\pi} d\sigma\; w_1 \eta_2=-\frac{c}{3\ell^2\ell_4^2}w_1 [\eta_2]_{zm}
\end{align}
where the only contribution comes from the winding terms. In terms of Fourier modes we can write:
\begin{align}
\{\bar{Q}_m,P_n\}&=-\frac{im k_4 p\lambda_{J\bar{T}}}{2 R_Q}\bar{Q}_m\delta_{n,0}
\end{align}

\subsubsection*{Right pseudoconformal-left conformal bracket $\{\bar{Q}_{\bar{f}_1},Q_{f_2}\}$}
In the case of the bracket between a right-moving pseudoconformal charge and a left-moving conformal one, the only non-trivial contribution comes from the winding terms:
\begin{align}
\{\bar{Q}_{\bar{f}_1},Q_{f_2}\}&=\delta_{f_2}\bar{Q}_{\bar{f}_1} =-\frac{ Q_L}{4\pi G_3 \ell_4}\int_0^{2\pi} w_1 f_2
\end{align}
For the Fourier modes we obtain:
\begin{align}
\{\bar{Q}_m,Q_n\}&=-\frac{im \lambda_{J\bar{T}}}{R_Q}\bar{Q}_m P_0 \delta_{n,0}
\end{align}
where, clearly, we can only see the order 0 charges $P_0 \delta_{n,0}$ appearing in the non-linear term at this order in the perturbative analysis.

\subsubsection*{Right pseudoconformal bracket $\{\bar{Q}_{\bar{f}_1},\bar{Q}_{\bar{f}_2}\}$}
We proceed to compute the Poisson bracket between two right-moving pseudoconformal charges:
\begin{align}
&\{\bar{Q}_{\bar{f}_1},\bar{Q}_{\bar{f}_2}\}=\frac{R_v}{4\pi G_3 \ell_4}\int_0^{2\pi} d\sigma\; (\bar{f}_2 \bar{f}'_1-\bar{f}_1 \bar{f}'_2)\bigg(\ell\ell_4(T_v^2+Q_R^2)+\ell\ell_4(T_v^2+Q_R^2)\bar{f}'_b+Q_R\bar{\eta}'_b\bigg)+\nonumber\\
&+\frac{Q_R}{4\pi G_3 \ell_4}\int_0^{2\pi}d\sigma\;(w_2 \bar{f}_1-w_1 \bar{f}_2)=\bar{Q}_{\bar{f}_1 \bar{f}'_2-\bar{f}_2 \bar{f}'_1}+\frac{c Q_R}{3 \ell\ell_4}\big(w_2 [\bar{f}_1]_{zm}-w_1 [\bar{f}_2]_{zm}\big)
\end{align}
We note that he terms linear in $\hat{v}$ canceled. In terms of Fourier modes we can write:
\begin{align}
\{\bar{Q}_m,\bar{Q}_n\}&=\frac{i (m-n)}{R_v}\bar{Q}_{m+n}+\frac{i m \lambda_{J\bar{T}}}{R_v R_Q}\bar{Q}_m\bar{P}_0 \delta_{n,0}-\frac{i n \lambda_{J\bar{T}}}{R_v R_Q}\bar{Q}_n\bar{P}_0 \delta_{m,0}
\end{align}

\subsubsection*{Right pseudoconformal-right affine bracket $\{\bar{Q}_{\bar{f}_1},\bar{P}_{\bar{\eta}_2}\}$}
Finally, the Poisson bracket between a right-moving pseudoconformal transformation and a right-moving affine transformation is given by:
\begin{align}
\{\bar{Q}_{\bar{f}_1},\bar{P}_{\bar{\eta}_2}\}&=-\frac{R_v}{4\pi G_3 \ell\ell_4^2}\int_0^{2\pi } d\sigma\; \bar{f}_1 \bar{\eta}'_2\bigg( \ell\ell_4 Q_R \bar{f}'_b+\bar{\eta}'_b\bigg)-\frac{1}{4\pi G_3 \ell\ell_4^2} \int_0^{2\pi } d\sigma\; \big( w_1 \bar{\eta}_2 - \ell\ell_4 Q_R w_2 \bar{f}_1\big)=\nonumber\\
&=\bar{P}_{\bar{f}_1\bar{\eta}'_2}-\frac{c}{3 \ell^2\ell_4^2}\big( w_1 [\bar{\eta}_2]_{zm} -\ell\ell_4 Q_R w_2 [\bar{f}_1]_{zm}\big)
\end{align}
For the Fourier modes, we obtain:
\begin{align}
\{\bar{Q}_m,\bar{P}_n\}&=-\frac{in \bar{P}_{m+n}}{R_v}-\frac{im p k_4 \lambda_{J\bar{T}}}{2R_Q}\bar{Q}_m \delta_{n,0}-\frac{i n \lambda_{J\bar{T}}}{R_v R_Q}\bar{P}_0\bar{P}_n\delta_{m,0}
\end{align}

\subsubsection*{Conclusion}
We computed perturbatively around constant backgrounds the asymptotic symmetry algebra in the charged warped BTZ phase space and we obtained a perfect match with the symmetry algebra of a symmetric product orbifold of $J\bar{T}$-deformed CFTs with $p$ copies, central charge of the seed $6k$ and level of the $U(1)$ Kac-Moody algebra of the seed $k_4$. In order to facilitate comparison, we listed the Poisson brackets of the $J\bar{T}$ symmetry algebra in appendix \ref{appB:JTbar}.

\section{Discussion and outlook}
\label{section4:discussion}
The holographic duals to warped AdS$_3$ and, more generally, warped BTZ backgrounds, are generally non-local field theories with global $SL(2,\mathbb{R})\times U(1)$ symmetry, features also shared by the holographic duals of (near-)extremal Kerr black holes. The asymptotic symmetries analysis of the near-horizon geometry of the latter reveals a symmetry enhancement to one copy of the Virasaro algebra on the ``non-local side".

From a field-theoretical perspective, it is not clear, in general, how to reconcile the Virasoro symmetry with non-locality. The $J\bar{T}$ deformation of two-dimensional CFTs provides explicit examples of non-local field theories with an infinite dimensional symmetry algebra, which takes the form of $(Virasoro\times U(1)Kac-Moody)^2$ in a particular basis for which the action of the generators is non-local (see \cite{Georgescu:2024ppd} for the case of a $J\bar{T}$-deformed free boson). In a different basis, in which the generators act more locally, in a sense discussed in \cite{Georgescu:2024ppd}, the Poisson algebra is non-linear. The fact that the $J\bar{T}$ deformation is highly tractable and leads to UV-complete theories can be interpreted as a consequence of the infinite dimensional symmetry algebra that constrains it. It is a natural question whether there exist other deformations of two-dimensional CFTs  whose structure is constrained by the same symmetries and whether we can relate them to Kerr holography, or to its simpler three-dimensional version, warped AdS$_3$ holography.

In this article, we showed that a particular warped BTZ background constructed in \cite{Apolo:2021wcn} in string theory provides a holographic realization of another such deformation. The asymptotic symmetry algebra that we obtained precisely reproduces the infinite dimensional $J\bar{T}$ algebra, in the  non-linear basis. In this set-up, the non-linearity can be interpreted as the bulk manifestation of the non-local nature of the dual non-gravitational theory. The distinctive features of the asymptotic symmetry analysis are absent if we turn off the $U(1)$ charges, which were not considered in early warped AdS$_3$ set-ups.

Our result strengthens the relation between this particular background obtained by TsT on $BTZ\times S^3\times T^4$ and the $J\bar{T}$ deformation. We interpret this relation, together with the existence of an infinite dimensional asymptotic symmetry algebra itself, as an indication that the deformation leads to a well-defined non-AdS holographic duality. It would be interesting to understand the non-gravitational theory, beyond the single-trace $J\bar{T}$ sector, and in particular to study the constraints imposed by the infinitely many conserved charges on the observables of the theory.

Deforming various holographic backgrounds by TsT often leads to interesting non-AdS holographic dualities. In the case of the $AdS_3\times S^3\times T^4$ background with pure RR flux which is obtained in the near-horizon of the D1-D5 system, the TsT deformation with one direction on AdS$_3$ and one of $S^3$ is known to correspond to a dipole deformation of the dual 2d CFT \cite{Song:2011sr}. In the case of the Poincare $AdS_3\times S^3\times T^4$ background with pure NS-NS flux, the TsT deformation with both directions on AdS$_3$ leads to the asymptotically linear dilaton background which is holographic dual to Little String Theory compactified to 2d \cite{Giveon:2017nie}. The same deformation applied to global $AdS_3\times S^3\times T^4$ was studied in \cite{Asrat:2023yzy}. For the background generated by the backreaction of a stack of D3-branes, the TsT deformation with both directions on the worldvolume leads, after taking the near-horizon limit, to the holographic dual of non-commutative SYM \cite{Maldacena:1999mh}. In \cite{Apolo:2021wcn}, the TsT deformation was applied, in particular, to $BTZ\times S^3\times T^4$ with pure NS-NS flux, with one direction on BTZ and one direction on $T^4$. On the non-gravitational side, it should correspond to an irrelevant deformation which breaks Lorentz symmetry and half of the conformal invariance. In general, it is not clear if such deformation leads to a well-defined, UV-complete theory. In the other examples mentioned, the non-AdS holographic dualities can be obtained directly from decoupling limits in string theory, which guarantees the UV-completeness of the non-gravitational theories. It would be interesting to understand if the warped BTZ background can also be obtained from a decoupling limit, which would also give information about its holographic dual, beyond the long string sector.

Since the warped BTZ background is obtained by applying the TsT transformation to the background which appears in the $\alpha'\rightarrow 0$ decoupling limit of the NS5-F1 system\footnote{Note that this is the S-dual of the near-horizon limit of the D1-D5 system and not the $g_s\rightarrow 0$ decoupling limit that leads to LST.}, a natural guess would be to apply the same transformation directly on the NS5-F1 system and then take the $\alpha'\rightarrow 0$ limit, with the appropriate scaling of the TsT deformation parameter in this limit. It is not hard to see that this naive attempt does not lead to the warped BTZ background that we considered, even in the simple case in which we set the $U(1)$ charges to 0. It remains an open question whether there exists a more complicated set-up for which a decoupling limit gives the warped BTZ background. Nevertheless, we note the difference with respect to the case of pure RR flux, for which the TsT and the near-horizon $\alpha'\rightarrow 0$ limit on the D1-D5 system commute.

Another reason why a decoupling limit of a brane system would be useful is that it might fix certain constants in the backgrounds, such as the asymptotic value of the ten-dimensional B-field. The thermodynamics is strongly affected by any constant shift in the  dilaton. It would be good if such constants, which are not seen at the level of the supergravity equations of motion, can be fixed via decoupling. In the similar, but better understood ``$T\bar{T}$" case, the dilaton which is obtained after the TsT transformation with both directions on AdS$_3$ also needs to be shifted in order for the electric charge to be quantized. However, by taking the $g_s\rightarrow 0$ decoupling limit of the NS5-F1 background, one can obtain directly the correct dilaton, which yields the quantized electric charge.

%indicating that the holographic dual theory is shares $J\bar{T}$-like features.
%A clear question is what is exactly is the holographic dual theory. For this reason, it would be useful to obtain the background directly from a decoupling limit in string theory, similar to how the linear dilaton background is obtained directly from the NS5-F1 decoupling limit. This should also fix the constant in the dilaton to that which was fixed in \cite{} by requiring conservation of the electric charge. This would give us an idea about the dual non-gravitational theory and about how far it is from single-trace $J\bar{T}$.

Regarding the technical details of our analysis, we set to 0 the radial functions of the allowed symmetry transformations, in order to secure the conservation and finiteness of the associated charges. It would be useful to look for counterterms which would allow for non-trivial radial functions, in particular fixed in terms of the rest of the functions, as it happens, for example, in the symmetry analysis of asymptotically AdS$_3$ spacetimes, resulting in central charge terms.

Since a worldsheet theory is available for the warped BTZ backgrounds supported by pure NS-NS flux analyzed in this article, it would be interesting to see if the method proposed in \cite{Du:2024tlu} to compute asymptotic symmetries directly from the worldsheet would reproduce our results. In the similar case related to the $T\bar{T}$ deformation presented in \cite{Du:2024bqk}, the full asymptotic symmetry transformations were fixed by requiring the equivalence of symplectic forms between the TsT string
theory and an auxiliary AdS$_3$ string theory. In particular, by imposing this constraint, one can reproduce holographically the central charge of the symmetry algebra of $T\bar{T}$-deformed CFTs. Thus, it would be worth checking if a similar requirement can fix the radial functions in our analysis, which we set to 0, to functions which would lead to the central charges of the Virasoro subalgebras of the $J\bar{T}$ symmetry algebra.

In the context of warped AdS$_3$ holography, a certain type of two-dimensional QFT, called Warped conformal field theory (WCFT) \cite{Hofman:2011zj,Detournay:2012pc}, was introduced as the non-gravitational dual to certain warped AdS$_3$ backgrounds. Warped CFTs have global $SL(2,\mathbb{R})\times U(1)$ symmetry, which gets enhanced to $(Virasoro\times U(1) Kac-Moody)$ and obey an universal entropy formula. There exists a redefinition of the symmetry generators that recasts this entropy formula into the standard Cardy formula. This redefinition, which is referred to as a change in ensemble from canonical to quadratic, is motivated by holographic realizations of WCFTs and it corresponds to a non-local modification of the WCFT symmetry algebra, for which the level of the $U(1)$ Kac-Moody algebra becomes charge dependent. The WCFT symmetries were reproduced holographically in both ensembles \cite{Aggarwal:2020igb}. Considering the similarity between this redefinition of generators and the one for $J\bar{T}$-deformed CFTs, it would be worth studying the precise relation, in an attempt to connect various top-down and bottom-up results in warped AdS$_3$ holography.

Finally, it would be interesting to study whether there exists a larger class of $U(1)$ charged backgrounds which share the same entropy formula and asymptotic symmetries as the symmetric product orbifold of $J\bar{T}$-deformed CFTs. If so, it would be worth understanding, more generally, what structure a background should have in order to exhibit these universal features. Inspired by the analysis of \cite{Georgescu:2024iam}, a natural starting point is the family of backgrounds obtained in \cite{El-Showk:2011euy} by using $(1,2)$ deformations of the D1-D5 system that generalize single-trace $J\bar{T}$, which should be extended by adding $U(1)$ charges.

%These backgrounds have the advantage of being obtained directly from decoupling limits and their non-gravitational duals are known to correspond to the low-energy limit of dipole deformations of the D1-D5 system. 

 \subsubsection*{Acknowledgements}
The author is grateful to Monica Guica for insightful discussions and comments on the draft. The work of SG is supported by the Royal Society-Newton International Fellowship NIF/R1/241888.

\appendix

\section{Covariant phase space formalism toolkit}
\label{appA:covphsp}
In this appendix, we list the expressions used in order to compute the various quantities from the main text, in the covariant phase space formalism, applied to the NS-NS sector of type IIB supergravity. Comprehensive reviews on the covariant phase space formalism can be found in \cite{Compere:2007vx,Compere:2009dp,Compere:2018aar}.

The NS-NS part of the type IIB sugra action is:
\begin{align}\label{typeiibaction}
S^{IIB}_{\text{NS-NS}}&=\frac{1}{16\pi G_{10}}\int d^{10}x \sqrt{-g}\bigg(R-\frac{1}{2}\nabla_{\mu}\Phi\nabla^{\mu}\Phi-\frac{1}{2}\frac{1}{3!}e^{-(\Phi-\Phi_0)}H_{\mu\nu\rho}H^{\mu\nu\rho}\bigg)
\end{align}
where the metric is in Einstein frame and $\Phi_0$ is the expectation value of the dilaton. In ten dimensions, the relation between the Einstein frame metric and the string frame metric is $g^{Ein}_{\mu\nu}=e^{-(\Phi-\Phi_0)/2}g^{str}_{\mu\nu}$.

The presymplectic potential corresponding to the action \eqref{typeiibaction}, evaluated on an arbitrary variation of the fields $\delta g_{\mu\nu},\delta B_{\mu\nu},\delta \Phi$, is given by:
\begin{align}
\Theta&=\frac{1}{16\pi G_{10}}(\Theta_g+\Theta_B+\Theta_{\Phi})
\end{align}
where the contributions from the NS-NS fields can be written as 
\begin{align}
\Theta^{\mu}_g&=\nabla_{\lambda}h^{\lambda\mu}-\nabla^{\mu}h\hspace{1cm}\Theta^{\mu}_B=-\frac{e^{-(\Phi-\Phi_0)}}{2}H^{\mu\alpha\beta}\delta B_{\alpha\beta}\hspace{1cm}\Theta^{\mu}_{\Phi}=-\nabla^{\mu}\Phi \delta\Phi
\end{align}
with $\Theta_{g,B,\Phi}=\frac{1}{9!}\Theta^{\mu}_{g,B,\Phi}\epsilon_{\mu\alpha_1...\alpha_9}dx^{\alpha_1}\wedge...\wedge dx^{\alpha_9}$. In the expressions above, $h_{\mu\nu} = \d g_{\mu\nu}$, $h=h_{\mu\nu}g^{\mu\nu}$, $\delta f^{\alpha_1 ...\alpha_n}=g^{\alpha_1\beta_1}...g^{\alpha_n\beta_n}\delta f_{\beta_1...\beta_n}$ and we assumed that $\delta\Phi_0=0$.

The presymplectic form, evaluated on two arbitrary variations of the fields labeled 1 and 2, can be similarly written as:
\begin{align}
\omega&=\frac{1}{16\pi G_{10}}(\omega_g+\omega_B+\omega_{\Phi})
\end{align}
where $\omega_{g,B,\Phi}=\frac{1}{9!}\omega^{\mu}_{g,B,\Phi}\epsilon_{\mu\alpha_1...\alpha_9}dx^{\alpha_1}\wedge...\wedge dx^{\alpha_9}$ has the various contributions given by:
\begin{align}
\omega^{\mu}_g&=\frac{1}{2}\bigg( (2\nabla_{\lambda}h_{1\rho}^{\mu}-\nabla^{\mu}h^1_{\lambda\rho})h_2^{\lambda\rho}-\nabla_{\rho}h_1 h_2^{\rho\mu}+h_1(\nabla_{\rho}h_2^{\rho\mu}-\nabla^{\mu}h_2)-(1\leftrightarrow 2)\bigg)\\
\omega^{\mu}_B&=-\frac{e^{-(\Phi-\Phi_0)}}{2}\bigg(\delta_1 H^{\mu\alpha\beta}\delta_2 B_{\alpha\beta}+\frac{h_1}{2}H^{\mu\alpha\beta}\delta_2 B_{\alpha\beta}-H^{\mu\alpha\beta}\delta_2 B_{\alpha\beta}\delta_1\Phi-(1\leftrightarrow 2)\bigg)\\
\omega^{\mu}_{\Phi}&=-\delta_1(\nabla^{\mu}\Phi)\delta_2\Phi - \frac{h_1}{2}\nabla^{\mu}\delta_2\Phi -(1\leftrightarrow 2)
\end{align}
For an asymptotic symmetry transformation corresponding to a diffeomorphism generated by $\xi$ and a B-field gauge transformation generated by $\Lambda$, the charge variation between two backgrounds which differ by $\delta g_{\mu\nu},\delta B_{\mu\nu},\delta\Phi$ is given by:
\begin{align}
\slash{\!\!\!\delta} Q_{\xi,\Lambda}&=\frac{1}{8\pi G_{10}}\frac{1}{2!8!}\int_{S^8}\epsilon_{\alpha_1\alpha_2...\alpha_8\mu\nu}k^{\mu\nu}dx^{\alpha_1}\wedge ...\wedge dx^{\alpha_8}
\end{align}
where $S^8$ is a closed spacelike surface and $k^{\mu\nu}=k^{\mu\nu}_{g,\xi}+k^{\mu\nu}_{B,\xi,\Lambda}+k^{\mu\nu}_{\Phi,\xi}$ with the following contributions:
\begin{align}
k^{\mu\nu}_{g,\xi}&= \frac{1}{2}\bigg(\xi^\nu  \nabla^\mu h - \xi^\nu \nabla_\s h^{\mu\s} + \xi_\s  \nabla^\nu h^{\mu\s} + \frac{1}{2} h  \nabla^\nu \xi^\mu - h^{\rho \nu} \nabla_\rho \xi^\mu - (\mu\leftrightarrow \nu)\bigg) \\
k^{\mu\nu}_{B,\xi,\Lambda}&=\frac{e^{-(\Phi-\Phi_0)}}{4}\bigg(H^{\mu\nu\lambda} (\xi^{\alpha}B_{\alpha\lambda}+\Lambda_{\lambda})\delta\Phi-H^{\mu\nu\lambda}\xi^{\alpha}\delta B_{\alpha\lambda}-\xi^{\mu}H^{\nu\alpha\lambda}\delta B_{\alpha\lambda}\nonumber\\
&+\big(2 h^{\mu\sigma}H_{\sigma}^{\hspace{0.2cm}\nu\lambda}+h^{\lambda\sigma}H^{\mu\nu}_{\hspace{0.4cm}\sigma}-\frac{h}{2} H^{\mu\nu\lambda}-\delta H^{\mu\nu\lambda}\big)(\xi^{\alpha}B_{\alpha\lambda}+\Lambda_{\lambda} )-(\mu\leftrightarrow\nu) \bigg)\\
k^{\mu\nu}_{\Phi,\xi}&=\frac{1}{2}(\xi^{\nu}\nabla^{\mu}\Phi-\xi^{\mu}\nabla^{\nu}\Phi)\delta\Phi
\end{align}
In our case, the dilaton is constant, so only the metric and the B-field contribute to the various quantities computed above.

\section{Perturbative analysis around constant backgrounds for double-trace $J\bar{T}$-deformed CFTs}
\label{appB:JTbar}
The aim of this appendix is to  particularize the results of \cite{Georgescu:2024ppd} to backgrounds with constant parameters and linearized perturbations around them, in order to facilitate the comparison between the warped $AdS_3$ analysis and the $J\bar{T}$ analysis.

A generic three-dimensional background analyzed in \cite{Georgescu:2024ppd} consists of an asymptotically $AdS_3$ metric and two $U(1)$ Chern-Simons gauge fields. For constant parameters, which we denote by $\mathcal{L}^{(0)}=T_u^2,\bar{\mathcal{L}}^{(0)}=T_v^2, \mathcal{J}^{(0)}=Q_L,\bar{\mathcal{J}}^{(0)}=Q_R$, it is given by:
\begin{align}
ds^2_3&=\ell^2 \, \frac{ dz^2}{z^2}+\bigg(T_u^2+T_v^2\frac{\l^2Q_L^2}{(1+\l Q_R)^2}-\frac{1}{z^2}(1+ z^4 T_u^2 T_v^2)\frac{\l Q_L}{1+\l Q_R}\bigg)dU^2+\nonumber\\
&+\frac{1}{1+\l Q_R}\bigg(\frac{1+ z^4 T_u^2 T_v^2}{z^2}-\frac{2\l T_v^2 Q_L}{1+\l Q_R}\bigg)dUdV+\frac{T_v^2}{(1+\l Q_R)^2}dV^2\\
A_U &= Q_L + \frac{\l^2 Q_L(\kappa T_v^2+Q_R^2)}{2(1+\l Q_R)}\;, \;\;\;\;\; A_V = -\frac{\l(\kappa T_v^2+Q_R^2)}{2(1+\l Q_R)}\;, \;\;\;\;\;\;\;\kappa\equiv \frac{k}{16\pi^2 G\ell} \nonumber \\
B_U &= -\frac{\l Q_L }{1+\l Q_R}\bigg[Q_R+\frac{\l(\kappa T_v^2+Q_R^2)}{2}\bigg] \;, \;\;\;\;\;\; B_V = \frac{Q_R}{1+\l Q_R}+ \frac{\l(\kappa T_v^2+Q_R^2)}{2(1+\l Q_R)}
\end{align}
We will use the index $(0)$ to denote the quantities corresponding to the background with constant parameters. The field-dependent coordinate and its radius are given by:
\begin{align}\label{vzeroorder}
v^{(0)}&=\frac{V-\lambda Q_L U}{1+\lambda Q_R}\hspace{1cm}R_v=\frac{1-\lambda Q_L}{1+\lambda Q_R}
\end{align}
The symmetry transformations in phase space found in \cite{Georgescu:2024ppd} are combinations of diffeomorphisms $\xi$ and gauge transformations $\Lambda_{A,B}$ for the two gauge fields. For constant parameters, they are given by:
\begin{align}
F_U&=f(U),\hspace{1cm}F_V=\bar{f}(v^{(0)})-\lambda(\eta(U)+\bar{\eta}(v^{(0)}))\label{symtrdiffeo}\\
\Lambda_A&=\eta(U)-\lambda Q_R (\bar{\eta}(v^{(0)})-\bar{\eta}_0)-\lambda (\kappa T_v^2+Q_R^2)(\bar{f}(v^{(0)})-\bar{f}_0)\\
\Lambda_B&=\bar{\eta}(v^{(0)})+\lambda Q_R (\bar{\eta}(v^{(0)})-\bar{\eta}_0)+\lambda (\kappa T_v^2+Q_R^2)(\bar{f}(v^{(0)})-\bar{f}_0)
\end{align}
where $\xi=F_U\partial_U+F_V\partial_V$ and $\bar{f}_0,\bar{\eta}_0$ are the zero modes of the corresponding functions. Compared to the analysis in \cite{Georgescu:2024ppd}, we set to 0 the radial component of the diffeomorphisms and ignored all the terms in the transformations that contain derivatives of the functions $f,\bar{f},\eta,\bar{\eta}$ higher than 1. We checked explicitly that by doing so, we only miss the Virasoro central charge term of the asymptotic symmetry algebra.

The four functions $f,\bar{f},\eta,\bar{\eta}$ that parametrize the symmetry transformations are periodic in their arguments. One can see explicitly that the winding terms in \cite{Georgescu:2024ppd} vanish on constant backgrounds.

The conserved charges corresponding to these symmetry transformations are all 0, except for the global ones, which correspond to the isometries $\partial_U,\partial_{-V}$ and to constant gauge transformations of the gauge fields
\begin{align}
E_L&=\frac{T_u^2}{8\pi G_{10}\ell} + \frac{2\pi}{k}Q_L^2,\hspace{1cm}E_R=\bigg(\frac{T_v^2}{8\pi G_{10}\ell}+\frac{2\pi}{k}Q_R^2\bigg)R_v\\
J_0&=Q_L-\frac{\lambda k}{4\pi}E_R,\hspace{1.8cm}\bar{J}_0=Q_R-\frac{\lambda k}{4\pi}E_R
\end{align}
The latter are the holonomies of the gauge fields around the boundary circle and correspond to the undeformed $U(1)$ charges in the $J\bar{T}$-deformed CFT.

Next, we want to generate a perturbed background by acting with a symmetry transformation in phase space parametrized by functions $f_b,\bar{f}_b,\eta_b,\bar{\eta}_b$. The new parameters are given by:
\begin{align}
\mathcal{L}=\mathcal{L}^{(0)}+\epsilon \mathcal{L}^{(1)} &=T_u^2\big(1+2\epsilon f_b'(U)\big),\hspace{1cm}\bar{\mathcal{L}}=\bar{\mathcal{L}}^{(0)}+\epsilon \bar{\mathcal{L}}^{(1)}=T_v^2\big(1+2\epsilon \bar{f}_b'(v^{(0)})\big)\nonumber\\
\mathcal{J}&=\mathcal{J}^{(0)}+\epsilon \mathcal{J}^{(1)}=Q_L +\epsilon\big(\eta'_b(U)+Q_L f'_b(U)\big)\nonumber\\
\bar{\mathcal{J}}&=\bar{\mathcal{J}}^{(0)}+\epsilon \bar{\mathcal{J}}^{(1)}=Q_R + \epsilon \big(\bar{\eta}_b'(v^{(0)})+Q_R \bar{f}'_b(v^{(0)})\big)
\end{align}
On the perturbed background, the field-dependent coordinate is $v=v^{(0)}+\epsilon v^{(1)}$ with:
\begin{align}
v^{(1)}&=-\frac{\lambda}{1+\lambda Q_R}\bigg(\eta_b(U)+\bar{\eta}_b(v^{(0)})+Q_L f_b(U)+Q_R \bar{f}_b(v^{(0)})\bigg)
\end{align}
For the symmetry transformations, it is clear that the $U$ component of the diffeomorphisms does not receive any correction. The $V$ component becomes $F_V=F_V^{(0)}+\epsilon F_V^{(1)}$ with $F_V^{(0)}$ in \eqref{symtrdiffeo} and
\begin{align}
F_V^{(1)}=-\frac{\lambda}{1+\lambda Q_R}\bigg(\eta_b(U)+\bar{\eta}_b(v^{(0)})+Q_L f_b(U)+Q_R \bar{f}_b(v^{(0)})\bigg)\big(\bar{f}'(v^{(0)})-\lambda \bar{\eta}'(v^{(0)})\big)
\end{align}
For the gauge transformations, we can similarly write:
\begin{align}
\Lambda_A^{(1)}+\Lambda_B^{(1)}&=\bar{\eta}'(v^{(0)})v^{(1)}\\
\Lambda_A^{(1)}-\Lambda_B^{(1)}&=-2\bigg(\lambda Q_R \bar{\eta}'(v^{(0)})+\frac{\lambda}{2}(\kappa T_v^2+Q_R^2)\bar{f}'(v^{(0)})\bigg)v^{(1)}-2\lambda \int^{v^{(0)}}\hspace{-0.4cm}\big(\bar{\eta}'_b(v^{(0)})+Q_R\bar{f}'_b(v^{(0)})\big)\bar{\eta}'(v^{(0)})-\nonumber\\
&-2\lambda\int^{v^{(0)}}\hspace{-0.4cm}\big((\kappa T_v^2+Q_R^2) \bar{f}'_b(v^{(0)})+Q_R\bar{\eta}'_b(v^{(0)})\big)\bar{f}'(v^{(0)})\label{primitives}
\end{align}
We chose to focus on these combinations because they appear in the warped AdS$_3$ analysis. On the perturbed background, we obtain four sets of non-trivial conserved charges, which are given by:
\begin{align}\label{conservedchJTbarlin}
P_{\eta}=\int_0^R d\sigma\; \eta(U)(\eta'_b(U)+Q_L f'_b(U))
\end{align}
\begin{align}
\bar{P}_{\bar{\eta}}&=-\frac{1-\lambda Q_L}{1+\lambda Q_R}\int_0^R d\sigma\; \bar{\eta}(v^{(0)}) (\bar{\eta}'_b(v^{(0)})+Q_R\bar{f}'_b(v^{(0)}))\\
Q_{f}&=\frac{4\pi}{k}\int_0^R d\sigma\; f(U) \big((\kappa T_u^2+Q_L^2)f'_b(U)+Q_L\eta'_b(U)\big) \\
\bar{Q}_{\bar{f}}&=-\frac{4\pi}{k}\frac{1-\lambda Q_L}{1+\lambda Q_R}\int_0^R d\sigma \; \bar{f}(v^{(0)})\big((\kappa T_v^2+Q_R^2) \bar{f}'_b(v^{(0)})+Q_R\bar{\eta}'_b(v^{(0)})\big)
\end{align}
Because of the primitives that appear in \eqref{primitives}, the corresponding transformations need to be supplemented by new ones that compensate for the winding terms, without affecting the conserved charges:
\begin{align}
\Lambda_A^{(1)w}&=w U,\hspace{1cm}\Lambda_B^{(1)w}=-w v^{(0)}
\end{align}
where the winding is given by:
\begin{align}
w&=\lambda \bigg[\big(\bar{\eta}'_b(v^{(0)})+Q_R\bar{f}'_b(v^{(0)})\big)\bar{\eta}'(v^{(0)})\bigg]_{zm}+\lambda\bigg[\big((\kappa T_v^2+Q_R^2) \bar{f}'_b(v^{(0)})+Q_R\bar{\eta}'_b(v^{(0)})\big)\bar{f}'(v^{(0)})\bigg]_{zm}=\nonumber\\
&=-\frac{1}{R_v}\bigg(\lambda \bar{P}_{\bar{\eta}'}+\frac{\lambda k}{4\pi}\bar{Q}_{\bar{f}'}\bigg)
\end{align}
Different from the notation in \cite{Georgescu:2024ppd}, the derivatives are with respect to $v^{(0)}$ and not $v^{(0)}/R_v$. Of course, we could have written also $\Lambda_A^{(1)w}=w v^{(0)}$. No extra terms are necessary in the diffeomorphisms, since $F_V$ is periodic.

Finally, we list here the Poisson brackets for the conserved charges, computed up to linear order around the constant backgrounds:
\begin{align}
\{P_m,P_n\}&=-\frac{im k}{2}\delta_{m+n,0},\hspace{1cm}\{P_m,\bar{P}_n\}=\frac{ink\lambda}{2 R_Q}\bar{P}_{n}\delta_{m,0}\\
\{\bar{P}_m,\bar{P}_n\}&=-\frac{i m k}{2}\delta_{m+n,0}+\frac{ink\lambda}{2R_Q}\bar{P}_n\delta_{m,0}-\frac{i m k\lambda}{2R_Q}\bar{P}_m\delta_{n,0}\\
\{Q_m,Q_n\}&=-i(m-n)Q_{m+n},\hspace{0.5cm}
\{Q_m,P_n\}=i n P_{m+n},\hspace{0.5cm}\{\bar{P}_{m},Q_{n}\}=-\frac{i m \lambda}{R_Q}\bar{P}_m P_0\delta_{n,0}\\
\{\bar{Q}_m,P_n\}&=-\frac{im k \lambda}{2 R_Q}\bar{Q}_m\delta_{n,0},\hspace{1cm}\{\bar{Q}_m,Q_n\}=-\frac{im \lambda}{R_Q}\bar{Q}_m P_0 \delta_{n,0}\\
\{\bar{Q}_m,\bar{Q}_n\}&=\frac{i (m-n)}{R_v}\bar{Q}_{m+n}+\frac{i m \lambda}{R_v R_Q}\bar{Q}_m\bar{P}_0 \delta_{n,0}-\frac{i n \lambda}{R_v R_Q}\bar{Q}_n\bar{P}_0 \delta_{m,0}\\
\{\bar{Q}_m,\bar{P}_n\}&=-\frac{in \bar{P}_{m+n}}{R_v}-\frac{im k\lambda}{2R_Q}\bar{Q}_m \delta_{n,0}-\frac{i n \lambda}{R_v R_Q}\bar{P}_0\bar{P}_n\delta_{m,0}
\end{align}

\section{Details of the allowed transformations on the perturbed backgrounds}
\label{appC:details}
In this appendix, we present the details of solving for the allowed transformations on the perturbed backgrounds. We list below the constraints that we obtain from the symplectic form, as explained in the main text:
\begin{align}
&\text{\textbullet} \hspace{0.2cm}\delta T_u :\hspace{0.5cm}\partial_V F_U^{(1)}=0 \label{ecper1}\\
&\text{\textbullet} \hspace{0.2cm}\delta T_v :\hspace{0.5cm}(\partial_U+\lambda Q_L\partial_V)\big(\ell \ell_4 F_V^{(1)}(U,V) - \lambda \Lambda_y^{(1)}(U,V)\big)=-\frac{\lambda}{1+\lambda Q_R}\bar{f}'(v)(\ell\ell_4 Q_L f'_b(U)+\eta'_b(U))\label{ecper2}
\end{align}
\begin{align}
&\text{\textbullet} \hspace{0.2cm}\delta Q_L :\hspace{0.5cm}\partial_V\bigg(-\ell_4^2 F_y^{(1)}(U,V)-\ell\ell_4\lambda (Q_R^2+T_v^2)F_V^{(1)}(U,V)+(1+2\lambda Q_R +\lambda^2 Q_R^2+\lambda^2 T_v^2)\Lambda_y^{(1)}(U,V)\bigg)-\nonumber\\
&\hspace{2cm}-2\ell\ell_4 Q_L(1+\lambda Q_R)\partial_V F_U^{(1)}(U,V)=\frac{2\lambda}{1+\lambda Q_R}\bar{\eta}'_b(v)\bigg( Q_R \bar{f}'(v) + \frac{\bar{\eta}'(v)}{\ell\ell_4}\bigg)+\nonumber\\
&\hspace{2cm}+\frac{2\ell\ell_4\lambda}{1+\lambda Q_R}\bar{f}'_b(v)\bigg( (T_v^2+Q_R^2)\bar{f}'(v)+\frac{Q_R}{\ell\ell_4}\bar{\eta}'(v) \bigg)\label{ecper3}\\
&\text{\textbullet} \hspace{0.2cm}\delta Q_R :\hspace{0.5cm}\ell\ell_4(-2Q_R -\lambda Q_R^2+\lambda T_v^2)(\partial_U+\lambda Q_L\partial_V)F_V^{(1)}(U,V)+\ell_4^2(\partial_U+\lambda Q_L\partial_V)F_y^{(1)}(U,V)+\nonumber\\
&\hspace{2cm}+(1+2\lambda Q_R +\lambda^2 Q_R^2-\lambda^2 T_v^2)(\partial_U+\lambda Q_L\partial_V)\Lambda_y^{(1)}(U,V)=\nonumber\\
&\hspace{2cm}=2\ell\ell_4\lambda\bigg(Q_L f'_b(U)+\frac{\eta'_b(U)}{\ell\ell_4}\bigg)\bigg(Q_R\bar{f}'(v)+\frac{\bar{\eta}'(v)}{\ell\ell_4}\bigg)\label{ecper4}
\end{align}
Since we are solving  equations involving only the derivatives of the functions that appear at linear order, we can in principle add extra constant transformations, expressed in terms of the parameters $T_{u,v},Q_{L,R}$ and/or the zero modes of the functions already introduced. We will ignore such constants in the following. Moreover, we note that the equations \eqref{ecper1}-\eqref{ecper4} do not depend on possible constants in the functions from order 0.

We start from \eqref{ecper1}, which imposes that $F_U^{(1)}$ is a function of $U$ only. Since the equation is identical to the one obtained on constant backgrounds, we set $F_U^{(1)}=0$, meaning that the function $f(U)$ does not receive any correction on the perturbed backgrounds.

Next, we solve \eqref{ecper2} for $F_V^{(1)}$, which gives:
\begin{align}\label{rezFvlin}
\ell\ell_4 F_V^{(1)}=\lambda \Lambda_y^{(1)}-\frac{\lambda}{1+\lambda Q_R}\bar{f}'(\ell\ell_4 Q_L (f_b(U)-f_{b0})+\eta_b(U)-\eta_{b0})+ g_v(v)
\end{align}
where $g_v(v)$ is a function that is not fixed for now, but should be determined in terms of the functions $f,\bar{f},\eta,\bar{\eta}$ from the zeroth order part of the transformations and the background functions. We denoted by $f_{b0},\eta_{b0}$ the zero modes of the functions $f_b(U),\eta_b(U)$, which we subtracted for later convenience. Of course, they can be absorbed in the definition of $g_v(v)$, as their coefficient is a function of $v$ only. Plugging in this solution in \eqref{ecper4}, we can write:
\begin{align}\label{rezFylin}
\ell_4^2 F_y^{(1)}+\Lambda_y^{(1)}&=\lambda (\ell\ell_4 Q_L (f_b(U)-f_{b0})+\eta_b(U)-\eta_{b0})\bigg( \frac{\lambda (T_v^2+Q_R^2)}{1+\lambda Q_R}\bar{f}'(v)+\frac{2}{\ell\ell_4}\bar{\eta}'(v)\bigg)+j_v(v)
\end{align}
where $j_v(v)$ is another arbitrary function of $v$ and we subtracted the same zero modes as above, which could have been absorbed in the definition of $j_v(v)$. Finally, extracting, for example, $F_y^{(1)}$ from \eqref{rezFylin} and plugging in \eqref{ecper3}, we find:
\begin{align}\label{sollambdalin}
&\Lambda_y^{(1)}=\frac{\lambda\bar{\eta}'(v)(\ell\ell_4 Q_L (f_b(U)-f_{b0})+\eta_b(U)-\eta_{b0})}{\ell\ell_4(1+\lambda Q_R)}+\frac{j_v(v)+\lambda(T_v^2+Q_R^2)g_v(v)}{2(1+\lambda Q_R)}+g_u(U)+\nonumber\\
&\hspace{-1cm}+\frac{1}{\ell\ell_4(1+\lambda Q_R)}\int^v dv'\bigg(\lambda \bar{\eta}'_b(v')(\ell\ell_4 Q_R\bar{f}'(v')+\bar{\eta}'(v'))+\ell\ell_4\lambda \bar{f}'_b(v')(\ell\ell_4(T_v^2+Q_R^2)\bar{f}'(v')+Q_R\bar{\eta}'(v'))\bigg)
\end{align}
where $g_u(U)$ is an arbitrary function of $U$. It is clear that the term on the second line is not a periodic function of $v$ and we can write its non-periodic part as:
\begin{align}
w_{\bar{f},\bar{\eta}} v:=&\frac{v R_v}{1+\lambda Q_R}\bigg( \frac{\lambda}{\ell\ell_4}[\bar{\eta}'(\ell\ell_4 Q_R \bar{f}'_b+\bar{\eta}'_b)]_{zm}+\lambda [\bar{f}'(\ell\ell_4(T_v^2+Q_R^2)\bar{f}'_b+Q_R \bar{\eta}'_b)]_{zm} \bigg)=\nonumber\\
&=-\frac{v}{1+\lambda Q_R}\frac{3\lambda}{c}\ell\ell_4(\bar{P}_{\bar{\eta}'}+\bar{Q}_{\bar{f}'})
\end{align} 
Such terms appear also in the components of the diffeomorphisms $F_V^{(1)}=\frac{\lambda}{\ell\ell_4 R_v}w_{\bar{f},\bar{\eta}}v+...$, $F_y^{(1)}=-\frac{1}{\ell_4^2 R_v}w_{\bar{f},\bar{\eta}}v+...$, where the dots represent periodic terms. Nevertheless, these functions need to be periodic in order to preserve the periodicities of the spacetime coordinates. The function $\Lambda_y^{(1)}$ also needs to be periodic, such that under B-field gauge transformations, the associated conserved charge, which is quantized, does not change. In order to cure this issue, let us first write down also the expressions that we obtain for $F_V^{(1)},F_y^{(1)}$, after plugging in the result for $\Lambda_y^{(1)}$. For simplicity, we will denote the second line in \eqref{sollambdalin} by $Pr(v)$, such that $Pr(v)=\frac{w_{\bar{f},\bar{\eta}}}{R_v}v+...$:
\begin{align}
F_V^{(1)}&=\frac{\lambda g_u}{\ell\ell_4}+\frac{\lambda(\ell\ell_4 Q_L (f_b-f_{b0})+\eta_b-\eta_{b0})(-\ell\ell_4\bar{f}'+\lambda\bar{\eta}')}{\ell^2\ell_4^2 (1+\lambda Q_R)}+\nonumber\\
&+\frac{1}{2\ell\ell_4(1+\lambda Q_R)}\bigg(\lambda j_v+(2+2\lambda Q_R +\lambda^2 T_v^2+\lambda^2 Q_R^2)g_v+2\lambda(1+\lambda Q_R)P_r\bigg)
\end{align}
\begin{align}
F_y^{(1)}&=-\frac{g_u}{\ell_4^2}+\frac{\lambda(\ell\ell_4 Q_L (f_b-f_{b0})+\eta_b-\eta_{b0})}{\ell\ell^3_4(1+\lambda Q_R)}\bigg(\ell\ell_4\lambda(T_v^2+Q_R^2)\bar{f}' +(1+2\lambda Q_R)\bar{\eta}' \bigg)+\nonumber\\
&+\frac{1}{2\ell_4^2(1+\lambda Q_R)}\bigg( -\lambda(T_v^2+Q_R^2)g_v +(1+2\lambda Q_R)j_v -2(1+\lambda Q_R)P_r \bigg)\\
\Lambda_y^{(1)}&=g_u + \frac{\lambda\bar{\eta}'(\ell\ell_4 Q_L (f_b-f_{b0})+\eta_b-\eta_{b0})}{\ell\ell_4(1+\lambda Q_R)}+\frac{1}{2(1+\lambda Q_R)}\bigg(j_v +\lambda(T_v^2+Q_R^2)g_v +2(1+\lambda Q_R)P_r \bigg)
\end{align}
We can use the arbitrary functions $g_v(v),j_v(v)$ which appeared from integrations in order to force the periodicity of at least two of the functions. We choose to write:
\begin{align}
&j_v+\lambda(T_v^2+Q_R^2)g_v+2(1+\lambda Q_R)P_r=J_v\\
&2(1+\lambda Q_R)g_v+\lambda J_v=G_v
\end{align}
with $J_v(v),G_v(v)$ are arbitrary periodic functions. Solving for $g_v,j_v$, we find that $F_V^{(1)},\Lambda_y^{(1)}$ are periodic, but $F_y^{(1)}$ is not:
\begin{align}
F_V^{(1)}&=\frac{\lambda g_u}{\ell\ell_4}+\frac{\lambda(\ell\ell_4 Q_L (f_b-f_{b0})+\eta_b-\eta_{b0})(-\ell\ell_4\bar{f}'+\lambda\bar{\eta}')}{\ell^2\ell_4^2 (1+\lambda Q_R)}+\frac{1}{2\ell\ell_4(1+\lambda Q_R)}G_v\\
F_y^{(1)}&=-\frac{g_u}{\ell_4^2}+\frac{\lambda(\ell\ell_4 Q_L (f_b-f_{b0})+\eta_b-\eta_{b0})}{\ell\ell^3_4(1+\lambda Q_R)}\bigg(\ell\ell_4\lambda(T_v^2+Q_R^2)\bar{f}' +(1+2\lambda Q_R)\bar{\eta}' \bigg)+\nonumber\\
&\hspace{-0.7cm}+\frac{1}{2\ell_4^2(1+\lambda Q_R)}\bigg((1+2\lambda Q_R+\lambda^2 T_v^2+\lambda^2 Q_R^2)J_v -\lambda(T_v^2+Q_R^2)G_v  -4(1+\lambda Q_R)^2 P_r \bigg)\\
\Lambda_y^{(1)}&=g_u + \frac{\lambda\bar{\eta}'(\ell\ell_4 Q_L (f_b-f_{b0})+\eta_b-\eta_{b0})}{\ell\ell_4(1+\lambda Q_R)}+\frac{1}{2(1+\lambda Q_R)}J_v
\end{align}
Hence, we need to add some compensating transformation, which is not visible on the constant backgrounds, in order to render $F_y^{(1)}$ periodic. Clearly, we can only add such compensating transformation if they are solutions of the homogeneous equations. By inspecting the general solution to the homogeneous equations \eqref{solutiontrct}, we find:
\begin{align}
\bar{f}=0\hspace{1cm}\eta=w_{\bar{f},\bar{\eta}} U\hspace{1cm}\bar{\eta}=-w_{\bar{f},\bar{\eta}}\frac{v}{R_v}
\end{align}
which corresponds to:
\begin{align}
F_y^{(1)w}=\frac{w_{\bar{f},\bar{\eta}}}{\ell_4^2}\bigg( U+(1+2\lambda Q_R)\frac{v}{R_v} \bigg)
\end{align}
and does not induce non-periodic terms to $F_V^{(1)},\Lambda_y^{(1)}$.

Next, we need to fix the functions $g_u(U),J_v(v),G_v(v)$, using the fact that $F^{(1)}_V,F^{(1)}_y,\Lambda^{(1)}_y$ should continue the functions $F^{(0)}_V,F^{(0)}_y,\Lambda^{(0)}_y$ on the perturbed backgrounds. By inspecting $\Lambda_y^{(0)}=-\eta(U)-\bar{\eta}(v)$, it is clear that the $\lambda$-independent function $\eta$ does not receive any correction, apart from the winding term $w_{\bar{f},\bar{\eta}}U$, which corresponds to setting $g_u(U)$ to 0. It is also clear that the continuation of $\bar{\eta}(v)$ should be of the form $\bar{\eta}'\delta v$, so it is convenient to write $J_v=\frac{2\lambda\bar{\eta}'}{\ell\ell_4}\tilde{J}_v$, such that:
\begin{align}\label{variationvv}
\delta v&=-\frac{\lambda}{\ell\ell_4(1+\lambda Q_R)}(\ell\ell_4 Q_L (f_b-f_{b0}) + \eta_b-\eta_{b0} +\tilde{J}_v)
\end{align}
Similarly, we should have:
\begin{align}
F_V^{(1)}&=\frac{1}{\ell\ell_4}(\ell\ell_4\bar{f}'-\lambda\bar{\eta}')\delta v
\end{align}
which determines $G_v$ in terms of $\tilde{J}_v$ as:
\begin{align}
G_v&=\frac{2\lambda}{\ell\ell_4}(-\ell\ell_4\bar{f}'+\lambda\bar{\eta}')\tilde{J}_v
\end{align}
This leads to the fact that we can write:
\begin{align}\label{fycontin}
F_y^{(1)}&=-\frac{2(1+\lambda Q_R)}{\ell_4^2}P_r +\frac{\lambda(\ell\ell_4 Q_L (f_b-f_{b0}) + \eta_b-\eta_{b0} +\tilde{J}_v)}{\ell_4^3\ell(1+\lambda Q_R)}(\ell\ell_4\lambda(T_v^2+Q_R^2)\bar{f}'+(1+2\lambda Q_R)\bar{\eta}')
\end{align}
Since $F_y^{(1)}=\bigg(-\frac{\lambda\ell(T_v^2+Q_R^2)}{\ell_4}\bar{f}'-\frac{1+2\lambda Q_R}{\ell_4^2}\bar{\eta}'\bigg)\delta v+...$, where the dots stand for the variation of the parameters $T_v,Q_R$ that appear in the expression, we need to identify this variation with the first term in \eqref{fycontin}. Writing $F_y^{(0)}=-\frac{1}{\ell^2}\int^v dv'(\lambda\ell\ell_4(T_v^2+Q_R^2)\bar{f}'(v')+(1+2\lambda Q_R)\bar{\eta}'(v'))+\frac{\eta(U)}{\ell^2_4}$, we obtain:
\begin{align}
\int^v \bigg(\lambda\ell\ell_4(T_v\delta T_v+Q_R\delta Q_R)\bar{f}'+\lambda\delta Q_R \bar{\eta}'\bigg)&=\frac{1}{\ell\ell_4}\int^v\bigg( \bar{\eta}'(\lambda\bar{\eta}'_b+\ell\ell_4\lambda Q_R\bar{f}'_b)+\nonumber\\
&+\bar{f}'(\lambda\bar{\eta}'_b\ell\ell_4 Q_R +\lambda\ell^2\ell^2_4(T_v^2+Q_R^2)\bar{f}'_b)\bigg) 
\end{align}
from which we identify:
\begin{align}\label{var1}
\delta Q_R&=\frac{\bar{\eta}'_b}{\ell\ell_4}+ Q_R \bar{f}'_b\hspace{1cm}\delta T_v^2=2T_v^2\bar{f}'_b
\end{align}
Next, we compute the variation of the field-dependent coordinate $v=\frac{V-\lambda Q_L U}{1+\lambda Q_R}$, which needs to take the form \eqref{variationvv}. It is clear that naively varying:
\begin{align}
\delta \bigg(\frac{V-\lambda Q_L U}{1+\lambda Q_R}\bigg)=-\frac{V-\lambda Q_L U}{(1+\lambda Q_R)^2}\lambda \delta Q_R-\frac{\lambda \delta Q_L U}{1+\lambda Q_R}
\end{align}
cannot lead to \eqref{variationvv}. Motivated by the analysis of \cite{Georgescu:2024ppd}, we write:
\begin{align}
v&=V-\lambda \int^U Q_L - \lambda \int^v Q_R 
\end{align}
such that the variation gives:
\begin{align}
\delta v&=-\frac{\lambda}{1+\lambda Q_R}\bigg(\int^U \delta Q_L+\int^v \delta Q_R\bigg)
\end{align}
Plugging in the expression for $\delta Q_R$, we find that:
\begin{align}\label{var2}
\delta Q_L&=\frac{\eta'_b}{\ell\ell_4}+Q_L f'_b\hspace{1cm}\tilde{J}_v=\ell\ell_4 Q_R (\bar{f}_b-\bar{f}_{b0})+\bar{\eta}_b-\bar{\eta}_{b0}+\mathcal{V}_0
\end{align}
where $\mathcal{V}_0$ is an integration constant. The zero modes $\bar{f}_{b0},\bar{\eta}_{b0}$ of the functions $\bar{f}_b(v),\bar{\eta}_b(v)$ can be absorbed into the definition of $\mathcal{V}_0$. For completeness, let us note that by comparing the variation of \eqref{resultzeroordch} and \eqref{conschargeor1}, we find:
\begin{align}\label{var3}
\delta T_u^2=2 T_u^2 f'_b
\end{align}
One can easily see that \eqref{var1} and \eqref{var2} are consistent with \eqref{resultzeroordch} and \eqref{conschargeor1}.

\section{The charged dipole background}
\label{appD:dipole}
In this appendix, we present a similar analysis for another $U(1)$ charged warped BTZ background, which leads to very different results compared to those from the main body of the paper. This warped BTZ background is also obtained by a TsT transformation on $BTZ\times S^3\times T^4$, but supported by pure RR flux instead of pure NS-NS flux. The TsT deformation in the bulk is known to map to a dipole deformation in the dual two-dimensional field theory, which renders it non-local. We will thus refer to this background as the charged dipole.

In the following, we set various constants to 1, as they do not influence the main features of the results that we are interested in\footnote{In particular, we set the Brown-Henneaux central charge $c=6$}. In order to construct the charged dipole background, we start from the near-horizon geometry of the non-extremal D1-D5 system, to which we add $U(1)$ charges:
\begin{align}\label{backgru1}
ds^2&=\frac{dr^2}{4(r^2-4T_u^2 T_v^2)}+rdUdV+T_u^2dU^2+T_v^2 dV^2+(dy+Q_L dU- Q_R dV)^2++d\Omega_3^2+\sum_{i=8}^{10} dy_i^2\nonumber\\
C_2&=\frac{\cos\theta}{4}d\phi\wedge d\chi +\frac{r}{2}dV\wedge dU+(-Q_L dU-Q_R dV)\wedge dy,\hspace{2cm}e^{\Phi}=1
\end{align}
All the other fields are 0. Next, we do a TsT transformation: T-duality on $y$, shift $V\rightarrow V+\gamma y$, T-duality back on $y$ and we obtain:
\begin{align}\label{backgrdipole}
ds^2&=\frac{dr^2}{4(r^2-4T_u^2 T_v^2)}+\frac{r}{1+\gamma^2 T_v^2}dUdV+\bigg(T_u^2- \frac{r^2\gamma^2}{4(1+\gamma^2 T_v^2)}\bigg)dU^2+\frac{T_v^2 }{1+\gamma^2 T_v^2}dV^2+\nonumber\\
&+\frac{(dy+Q_L dU- Q_R dV)^2}{1+\gamma^2 T_v^2}+d\Omega_3^2+\sum_{i=8}^{10}dy_i^2,\hspace{0.7cm}e^{2\Phi}=\frac{1}{1+\gamma^2 T_v^2}\\
B&=\frac{\gamma r}{2(1+\gamma^2 T_v^2)}dy\wedge dU + \frac{\gamma T_v^2}{1+\gamma^2 T_v^2}dy\wedge dV -\frac{\gamma r Q_R +2 Q_L \gamma T_v^2}{2(1+\gamma^2 T_v^2)}dV\wedge dU\nonumber\\
F_1&=0,\hspace{0.6cm}F_3=\frac{\sin\theta}{4}d\theta\wedge d\chi\wedge d\phi+\frac{1}{2}dr\wedge dV\wedge dU\nonumber\\
F_5&=\frac{\gamma}{2}dr\wedge dU\wedge \omega_{T^3}+\frac{\gamma\sin\theta}{8(1+\gamma^2 T_v^2)}\bigg[(r Q_R + 2 Q_L T_v^2)dU\wedge dV -2 T_v^2 dV\wedge dy -\nonumber\\
&- r dU\wedge dy\bigg]\wedge d\theta\wedge d\chi\wedge d\phi\nonumber
\end{align}
where $\omega_{T^3}=dy_8\wedge dy_9\wedge dy_{10}$. Different from the case with pure NS-NS flux, it is easy to check that the TsT transformation and the near-horizon limit commute for the D1-D5 system. Hence, this warped background is obtained directly from a decoupling limit of a brane system.

We are interested in computing the various global charges for this background. They generally receive contributions from all the fields and, in particular, depend on constant terms in the potentials. For this reason, we list also the RR potentials obtained by TsT:
\begin{align}
C_2&=\frac{\cos\theta}{4}d\phi\wedge d\chi+\bigg[\frac{r}{2}+\frac{\gamma^2 Q_R}{1+\gamma^2 T_v^2}\bigg(Q_L T_v^2+\frac{r Q_R}{2}\bigg)\bigg]dV\wedge dU -\frac{Q_R}{1+\gamma^2 T_v^2}dV\wedge dy+\nonumber\\
&+\bigg(\frac{\gamma^2 r Q_R}{2(1+\gamma^2 T_v^2)}-Q_L\bigg)dU\wedge dy,\hspace{0.6cm}C_0=-\gamma Q_R\\
C_4&=\gamma\bigg(\frac{r}{2}dU+T_v^2 dV\bigg)\wedge \omega_{T^3} +\frac{\gamma\cos\theta}{8(1+\gamma^2 T_v^2)}d\phi\wedge d\chi\wedge\bigg[(r Q_R + 2 Q_L T_v^2)dU\wedge dV -\nonumber\\
&-2 T_v^2 dV\wedge dy - r dU\wedge dy\bigg]\nonumber
\end{align}
They are related to the field-strength by 
\begin{align}
    F_1=dC_0,\hspace{1cm}F_3=dC_2-C_0 H,\hspace{1cm}F_5=dC_4-H\wedge C_2
\end{align}
The expressions for the conserved charges in type IIB supergravity were derived in the covariant phase space formalism in \cite{Dias:2019wof}. First, we are interested in the conserved charges associated to the isometries $\partial_U,\partial_{-V},\partial_y$. We already listed the contributions from the metric and dilaton to the charge variation in appendix \ref{appA:covphsp}. The axion does not contribute since $F_1=0$. The remaining contributions from the form fields to the variation in phase space of the charge $Q_{\xi}$ corresponding to an isometry $\xi$ are given by:
\begin{align}\label{generaltypeiib}
&\slash{\!\!\!\delta} Q_{\xi}=\frac{1}{32\pi G_{10} }\int_{S^8}\bigg[\delta\bigg((e^{\Phi}*F_3-C_4\wedge H)\wedge i_{\xi}C_2+(e^{-\Phi}*H-e^{\Phi}C_0 *F_3+C_4\wedge dC_2-\nonumber\\
&-F_5\wedge C_2)\wedge i_{\xi}B+F_5\wedge i_{\xi}C_4\bigg)+i_{\xi}\bigg(\delta C_2\wedge (e^{\Phi}*F_3-C_4\wedge H)+\delta B\wedge (e^{-\Phi}*H-\nonumber\\
&-e^{\Phi}C_0 *F_3+C_4\wedge dC_2-F_5\wedge C_2)+\delta C_4\wedge F_5\bigg)\bigg]
\end{align}
Computing explicitly the charge variations for $\partial_U,\partial_{-V}$ for the charged dipole background \eqref{backgrdipole}, we obtain divergent expressions. However, we can render the expressions finite by shifting the axion to $C_0=0$. In order to keep $F_3$ fixed, we need to shift also $C_2\rightarrow C_2-C_0 B$, which gives the simple expression:
\begin{align}
C_0&=0,\hspace{0.6cm}C_2=\frac{\cos\theta}{4}d\phi\wedge d\chi +\frac{r}{2}dV\wedge dU+Q_L dy\wedge dU+ Q_R dy\wedge dV
\end{align}
With this modification, the divergences in the metric contribution are canceled by the contributions from the form fields \eqref{generaltypeiib}. The final results take a very simple form:
\begin{align}
E_L&:=Q_{\partial_U}=T_u^2+Q_L^2,\hspace{1cm}E_R:=Q_{-\partial_V}=T_v^2+Q_R^2,\hspace{1cm}Q_{\partial_y}=Q_L+Q_R
\end{align}
Next, we compute the charge associated to gauge transformations of the RR 2-form field generated by $\Lambda=dy$\footnote{The full symmetry transformation is $C_2\rightarrow C_2+d\Lambda$, $C_4\rightarrow C_4-B\wedge d\Lambda$.} and obtain:
\begin{align}
Q_{dy}&=Q_R-Q_L
\end{align}
Different from the results for \eqref{backgrwarp}, the global charges of \eqref{backgrdipole} do not depend on the TsT deformation parameter $\gamma$. We can define the combinations:
\begin{align}
\hat{Q}_L&=\frac{Q_{\partial_y}-Q_{dy}}{2}=Q_L,\hspace{1cm}\hat{Q}_R=\frac{Q_{\partial_y}+Q_{dy}}{2}=Q_R
\end{align}
The Bekenstein-Hawking entropy is given by:
\begin{align}
S&=2\pi \big( T_u+T_v\big)
\end{align}
which we can express in terms of the global charges as:
\begin{align}
S&=2\pi\bigg(\sqrt{E_L-\hat{Q}_L^2}+\sqrt{E_R-\hat{Q}_R^2}\bigg)
\end{align}
The result is identical to the CFT one obtained for the undeformed background \eqref{backgru1} and clearly different from the $J\bar{T}$ result \eqref{entropynsns} obtained for the warped BTZ background supported by pure NS-NS flux \eqref{backgrwarp}.

We are also interested in the asymptotic symmetry transformations. Using the same method outlined in section \ref{section3:asysym}, we evaluate the symplectic form in phase space on a variation generated by \eqref{trstart} and   $\delta=\delta T_u\partial_{T_u}+\delta T_v\partial_{T_v}+\delta Q_L\partial_{Q_L}+\delta Q_R\partial_{Q_R}$ and impose that it vanishes asymptotically. We emphasize the fact that now the symplectic form receives contributions also from the RR fields. The general expression is given by:
\begin{align}
\omega(\delta_1,\delta_2)&=\frac{1}{32\pi G_{10}}\bigg[\delta_2 C_0 \delta_1 (2 e^{2\Phi}* F_1)+\delta_2 C_2\wedge(e^{\Phi}*F_3-C_4\wedge H)+\delta_2 C_4\wedge \delta *F_5+\nonumber\\
&+\delta_2 B\wedge(e^{-\Phi}*H-e^{\Phi}C_0 *F_3+C_4\wedge dC_2-F_5\wedge C_2) -(1\leftrightarrow 2)\bigg]
\end{align}
The remaining contributions from the metric and dilaton are listed in appendix \ref{appA:covphsp}. We obtain the following equations for the allowed transformations:
\begin{align}\label{eqdipolew}
&\text{\textbullet} \hspace{0.2cm}\delta T_u :\hspace{0.5cm}\partial_V F_U(U,V)=0\hspace{8cm}\\
&\text{\textbullet} \hspace{0.2cm}\delta T_v :\hspace{0.5cm}\partial_U F_V(U,V)=0\\
&\text{\textbullet} \hspace{0.2cm}\delta Q_L :\hspace{0.5cm} 2Q_L \partial_V F_U(U,V)+\partial_V\big(F_y(U,V)-\Lambda_y(U,V)\big)=0\\
&\text{\textbullet} \hspace{0.2cm}\delta Q_R :\hspace{0.5cm}2Q_R \partial_U F_V(U,V)-\partial_U\big(F_y(U,V)+\Lambda_y(U,V)\big)=0
\end{align}
The solution is very simple:
\begin{align}
\xi&=f(U)\partial_U +\bar{f}(V)\partial_V+\big(\eta(U)-\bar{\eta}(V)\big)\partial_y,\hspace{0.5cm}\Lambda=-\big(\eta(U)+\bar{\eta}(V)\big)dy
\end{align}
We note that there is no dependence on the TsT deformation parameter, no emergence of a field-dependent coordinate similar to \eqref{fielddepcoo0} and, in fact, no field-dependence at all in the allowed transformations. Perturbing the background with such an allowed transformation and redoing the analysis of section \ref{section33:allowedperturbed}, we obtain the same trivial equations. In particular, there are no charge-dependent parameters entering the symmetry transformations and hence no non-linear terms in the asymptotic symmetry algebra.

The main point that we wanted to emphasize in this appendix is that for a similar charged warped BTZ background, the thermodynamics and asymptotic symmetries seem to be very different from those obtained for \eqref{backgrwarp}. In particular, the non-locality of the dual theory is not visible from the symmetry transformations alone, or at least not in the same way as for \eqref{backgrwarp}. We do not rule out the possibility that there exist modifications of this set-up (for example adding different $U(1)$ charges to the near-horizon of the D1-D5 system) that might reproduce the $J\bar{T}$-like features of \eqref{backgrwarp}. Another possibility is that the charged dipole background reproduces the $J\bar{T}$ symmetry algebra in the basis in which it is linear and the non-locality is hidden in the generators, although they do not depend on any field-dependent coordinate. Regardless, it would be worth pursuing a unified picture of (asymptotic) symmetries within warped AdS$_3$ holography, especially in examples provided by string constructions.

\end{document}